\documentclass[review]{elsarticle}

\usepackage{hyperref}
\usepackage{graphicx}
\usepackage{caption}
\usepackage{subcaption}
\usepackage{textgreek}
\usepackage{xcolor}
\definecolor{cinnamon}{rgb}{0.82, 0.41, 0.12}
\definecolor{pink}{rgb}{0.858, 0.188, 0.478}
\definecolor{black}{rgb}{0.0, 0.0, 0.0}

\def\sz#1{{\textcolor{black}{#1}}}

\usepackage[utf8]{inputenc}

\journal{Journal of \LaTeX\ Templates}

\graphicspath{{Figures/}}

\begin{document}

\begin{frontmatter}

\title{Turbulence modulation by finite-size spherical particles in Newtonian and viscoelastic fluids}

%% Group authors per affiliation:
\author[mymainaddress]{Sagar Zade\corref{mycorrespondingauthor}}
\cortext[mycorrespondingauthor]{Corresponding author}
\ead{zade@mech.kth.se}

\author[mymainaddress]{Fredrik Lundell} 

\author[mymainaddress]{Luca Brandt}

\address[mymainaddress]{Linn\'e Flow Centre and SeRC (Swedish e-Science Research Centre), \\ KTH Mechanics, SE 100 44 Stockholm, Sweden \\ Tel.: +46 727746122 \\}
%% or include affiliations in footnotes:

\begin{abstract}
We experimentally investigate the influence of finite-size spherical particles in turbulent flows of a Newtonian and a drag reducing viscoelastic fluid at varying particle volume fractions and fixed Reynolds number. Experiments are performed in a square duct at a Reynolds number $Re_{2H}$ of nearly $1.1\times10^4$, Weissenberg number $Wi$ for single phase flow is between 1--2 and results in a drag-reduction of 43\% compared to a Newtonian flow (at the same $Re_{2H}$). Particles are almost neutrally-buoyant hydrogel spheres having a density ratio of 1.0035$\pm$0.0003 and a duct height ${2H}$ to particle diameter $d_p$ ratio of around 10. We measure flow statistics for four different volume fractions $\phi$ namely 5, 10, 15 and 20\% by using refractive-index-matched Particle Image Velocimetry (PIV). For both Newtonian Fluid (NF) and Viscoelastic Fluid (VEF), the drag monotonically increases with $\phi$. For NF, the magnitude of drag increase due to particle addition can be reasonably estimated using a concentration dependent effective viscosity for volume fractions below 10\%. The drag increase is, however, underestimated at higher $\phi$. For VEF, the absolute value of drag is lower than NF but, its rate of increase with $\phi$ is higher. Similar to particles in a NF, particles in VEF tend to migrate towards the center of the duct and form a layer of high concentration at the wall. Interestingly, relatively higher migration towards the center and lower migration towards the walls is observed for VEF. The primary Reynolds shear stress reduces with increasing $\phi$ throughout the duct height for both types of fluid.
\end{abstract}

%\begin{keyword}
%
%\end{keyword}

\end{frontmatter}

% \linenumbers

\section{Introduction}

Turbulent flow of suspensions is encountered in many natural situations e.g.\ transport of sediments, flow of red blood cells in the body, etc.\ and industrial applications e.g.\ transport of crushed coal, slurries, particle dispersions in paints, foodstuffs, etc. This article focuses on a suspension of spherical particles in a square duct. Turbulent characteristics of the single phase flow are modified to varying extents based on particle size \citep{costa2017finite}, shape \citep{ardekani2017sedimentation}, concentration \citep{lashgari2014laminar}, density ratio \citep{fornari2016effect} and deformability \citep{alizad2017numerical}. Velocity and particle distribution determine the friction at the wall which is of fundamental importance in estimating power consumption in process industries. 

Suspension of finite-size particles in wall-bounded Newtonian flows have been shown to exhibit a variety of rich physics. In the viscous Stokes regime, particles migrate from regions of high shear to low shear due to irreversible interactions, e.g. towards the centerline in a Poiseuille flow \citep{guazzelli2011physical}. With an increase in particle $Re$, inertial effects become important and particles tend to move away from the centerline and equilibrate at an intermediate position due to the repulsive forces from the wall (see the tubular pinch effect in \citet{segre1962behaviour}). Dilute laminar flow of finite-size particles is known to exhibit an increase in the effective viscosity \citep{guazzelli2011physical}. \citet{bagnold1954experiments} showed how inter-particle collisions increase the effective viscosity in the highly inertial regime. Such inertial effects at the particle scale can induce other rheological effects like shear-thickening \citep{picano2013shear}. \citet{lashgari2014laminar} showed how the distribution of viscous, turbulent and particle stresses varies when changing the particle volume fraction and the $Re$ in a plane channel flow. For a square duct, \citet{kazerooni2017inertial} studied numerically the suspension of laminar flow at different $Re$, up to a particle volume fraction $\phi$ = 20\%, and for different duct to particle size ratios. According to their study, particles largely move to the corners at lower volume fractions and at higher $Re$. \citet{fornari2017suspensions} investigated turbulent flows of a suspension of spherical rigid particles in a square duct up to a volume fraction of 20\% and found that at the highest volume fraction, particles preferentially accumulate in the core region and the intensity of the secondary flows reduces below that of the unladen case.

It is well known that addition of trace amounts, e.g. few parts per million, of long-chain polymer in to a (soluble) solution leads to a remarkable decrease in the wall friction, referred to as Tom's effect \citep{toms1948some}. This drag reduction capability has been successfully used in crude-oil pipelines for increasing the flow rate at fixed pumping costs, the most famous example being the Trans Alaskan Pipeline in 1979 \citep{burger1980studies}, in preventing flooding by increasing the discharge of sewage during excessive rainfall \citep{sellin1980polymer}, district heating and cooling \citep{leca1984drag}, etc. Polymer additives are particularly attractive for industrial applications since only minute quantities can have substantial drag-reducing effect. 

These high molecular polymers dissolve in the solvent liquid and form coiled microstructures that have elastic properties and thus the resulting solution is viscoelastic in its rheology. When the relaxation time $\lambda$ of these microstructures is comparable or larger than the characteristic deformation time of the flow  $1/\dot{\epsilon}$, $\dot{\epsilon}$ being the extensional strain rate, these coiled microstructures stretch and substantially increase the elongational viscosity of the solution. The increased elongational viscosity, which mostly occurs in the near-wall region, where $\dot{\epsilon}$ is the highest, suppresses turbulent fluctuations. The effectiveness of polymer solutions, thus, depends on the stretching of individual molecules by the stresses in the flow \citep{gyr2013drag}. The Weissenberg number $Wi$, given by $\lambda \dot{\epsilon}$, compares the elastic forces to the viscous forces in the fluid. 

With increasing drag reduction, there is an increase in the spanwise spacing between the low-speed velocity streaks and there is a reduction in the number and strength of near-wall vortical structures while their size also increases \cite{white2004turbulence}.
Turbulence is attenuated at small scales due to the increasing elastic energy stored in the stretched coils at the small scales (\sz{owing to higher stretching dynamics at small scales}), thus interfering with the usual turbulence cascade mechanism \citep{sreenivasan2000onset}. Reynolds shear stress is substantially reduced leading to a reduction in cross-stream momentum transfer. The drag reduction is ultimately bounded by the maximum drag reduction asymptote \citep{virk1975drag} where the Reynolds shear stress reduces to nearly zero but, turbulence is sustained because of the interaction between fluctuating polymer stresses and the fluctuating velocity gradient \citep{warholic1999influence}. Also see \citet{hara2017experimental}, who experimentally studied the Reynolds number dependency of this interaction term. Amongst the many proposed mechanisms for regeneration of polymer wall turbulence, \citet{dubief2004coherent} found that polymer chains extract energy from the near-wall vortices ($y^+\geq$20) as they are pulled around the vortices, and release energy in the high speed streaks that are located just above the viscous sublayer ($y^+\approx$ 5) thus, causing an autonomous regeneration cycle.

Regarding VEF flow in a square duct, \citet{gampert1996polymer} found that with increasing polymer concentration in a square duct, the axial turbulence intensity first increases and then decreases even below the level obtained with a pure solvent. \citet{escudier2001fully} performed detailed spatial measurements of mean axial and secondary flow velocity as well as turbulence statistics for various polymer solutions in the duct. They found that apart from a reduction in the transverse turbulence intensity, there is also a strong reduction in the secondary flow velocities. \citet{owolabi2017turbulent} measured drag reduction in turbulent flow through ducts of various cross-sections and at varying degrees of mechanical degradation of polymer molecules. They found that the drag reduction, at least for flexible linear polymer additives, is a function of the $Wi$ (estimated using the fluid relaxation time $\lambda$ and the mean shear rate at the wall) only. \sz{\citet{Shahmardi1197576} performed direct numerical simulations (DNS), using the FENE-P model, to study the modulation of secondary flow. They found that, compared to NF case, the counter rotating vortices become larger and their centers are displaced towards the center of the duct away from the walls.}

\sz{In many industrial processes, e.g. food-processing, particles are suspended in a VEF medium.} In addition, considering the high effectiveness of drag reducing polymer additives in single phase flow, it is of practical importance to assess their effectiveness in a suspension flow. However, studies in this field mostly deal with the motion of inertia-less particles passively transported in VEF \sz{(see \cite{nowbahar2013turbophoresis} and review by \cite{d2015particle})}. For finite-size particles in VEF, very few studies exists and even those are mostly related to motion of a single particle at low $Re$. \citet{van1993effects} found that settling velocity of a spherical particle is reduced by elastic effects (e.g.\ presence of normal stress differences and high elongational viscosity) in the fluid, and that this effect becomes significantly higher with increasing shear rates experienced by the falling sphere. \citet{michele1977alignment} observed alignment and aggregation of spheres in plane shear flows. \citet{li2015dynamics} numerically studied the migration of a sphere in laminar square duct flow and found that the equilibrium position depends on the interplay between the elastic (driving the particle towards the channel center line) and inertial effects (drives the particle away from the channel center line). Also shear-thinning effects and secondary flows tend to move the particle away from the channel center line. Dramatic reduction in particle mobility, i.e.\ the tensor of proportionality between applied force and particle velocity, is seen due to viscoelastic wake structures, that are linked to an increase in the form drag \citep{murch2017growth}. Recently, \citet{einarsson2018einstein} analytically calculated the suspension stress for a dilute suspension of spheres in a viscoelastic medium and showed how shear-thickening arises from strain `hot spots' in the disturbance flow around particles.

Most of the studies in suspension flows are performed using numerical tools because of its ability to provide spatiotemporally resolved data which can shed light on the phyiscal mechanism behind the bulk observables. High performance computers have made possible fully resolved DNS at moderate Reynolds number. Using Immersed Boundary Method (IBM) of forcing, where the particle geometry is resolved, it has been possible to capture fluid-particle interaction with high fidelity \citep{breugem2012second}. However, to our knowledge, such a coupling has not been extended to suspension of turbulent viscoelastic flow. Experiments, on the other hand, are quite challenging due to lack of a convenient measurement techniques where both time and spatially resolved data can be acquired. Optical techniques fail with opaque particles where it becomes impossible to \textit{see} in the core of the flow. Refractive index matched particles have been used with quite some success to overcome this hurdle \citep{wiederseiner2011refractive, byron2013refractive, klein2012simultaneous, zade2018experimental} and have been used in this study.

\subsection{Outline}

In this study, we perform experiments with a suspension of spherical particles in a Newtonian fluid (NF) as well as a drag reducing viscoelastic fluid (VEF). The wall-bounded geometry is a square duct and measurements are performed in the center-plane i.e.\ the plane of the wall-bisector. In order to make a consistent comparison for both types of fluid flows, the Reynolds number is kept constant $Re_{2H} = 1.1\times10^4$. We study the change in wall friction and turbulent velocity statistics as a function of the particle volume fraction. By being able to differentiate between the fluid and particle phase, it is also possible to measure the mean particle concentration and velocity profile in the measurement plane.

In the following sections we first describe the experimental set-up and the measurement techniques along with information pertaining to particles and the rheology of the VEF. Later we present results, first with particles in NF and later in VEF. Finally we conclude by comparing the most interesting differences between suspension in the two fluid types.

\section{Experimental technique}

\subsection{Experimental set-up}

\begin{figure}
\centering
\begin{subfigure}{.45\textwidth}
  \centering
  \includegraphics[height=0.85\linewidth]{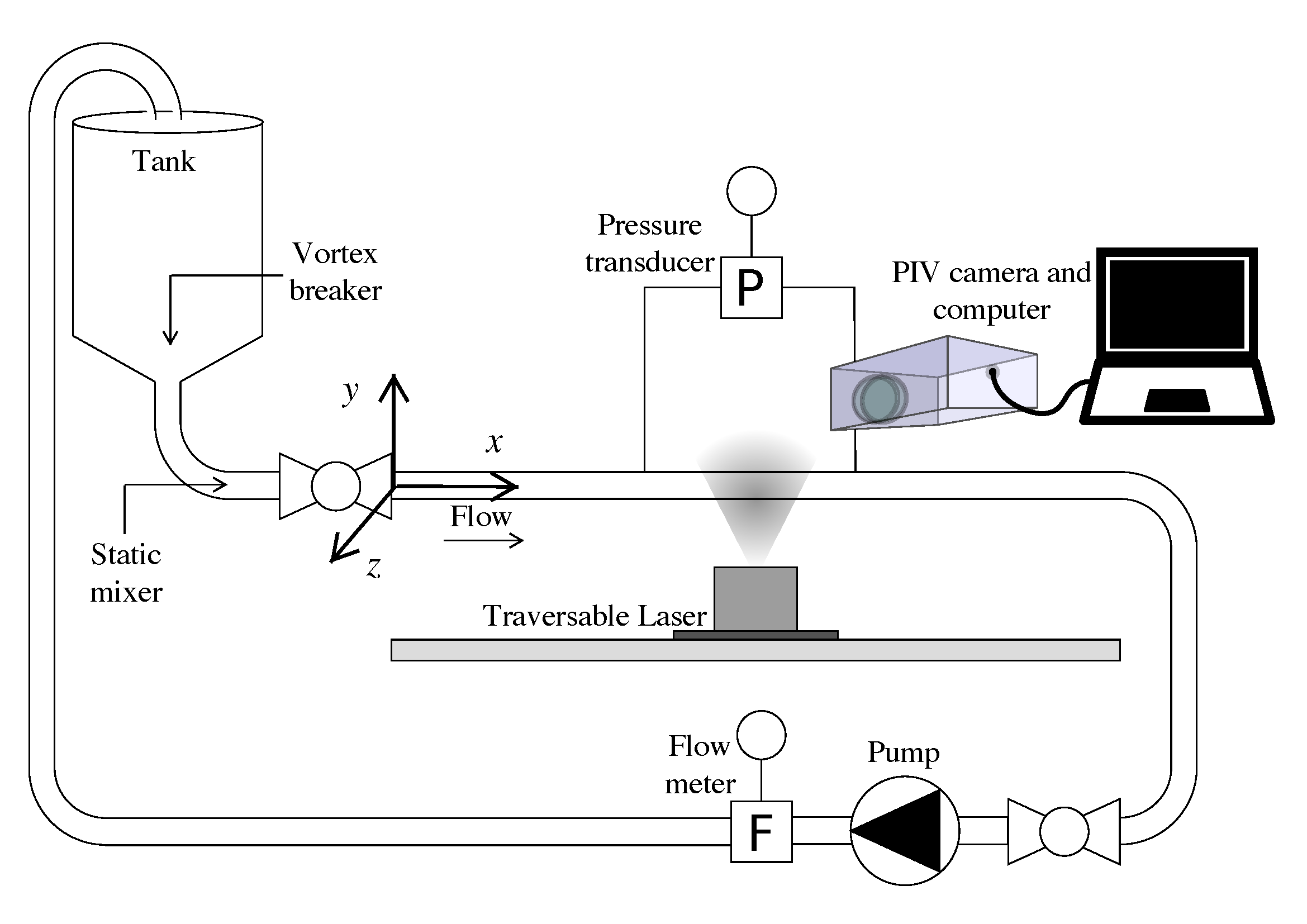}
  \caption{}
  \label{fig:Set-up schematic}
\end{subfigure}%
\begin{subfigure}{.45\textwidth}
  \centering
  \includegraphics[height=0.75\linewidth]{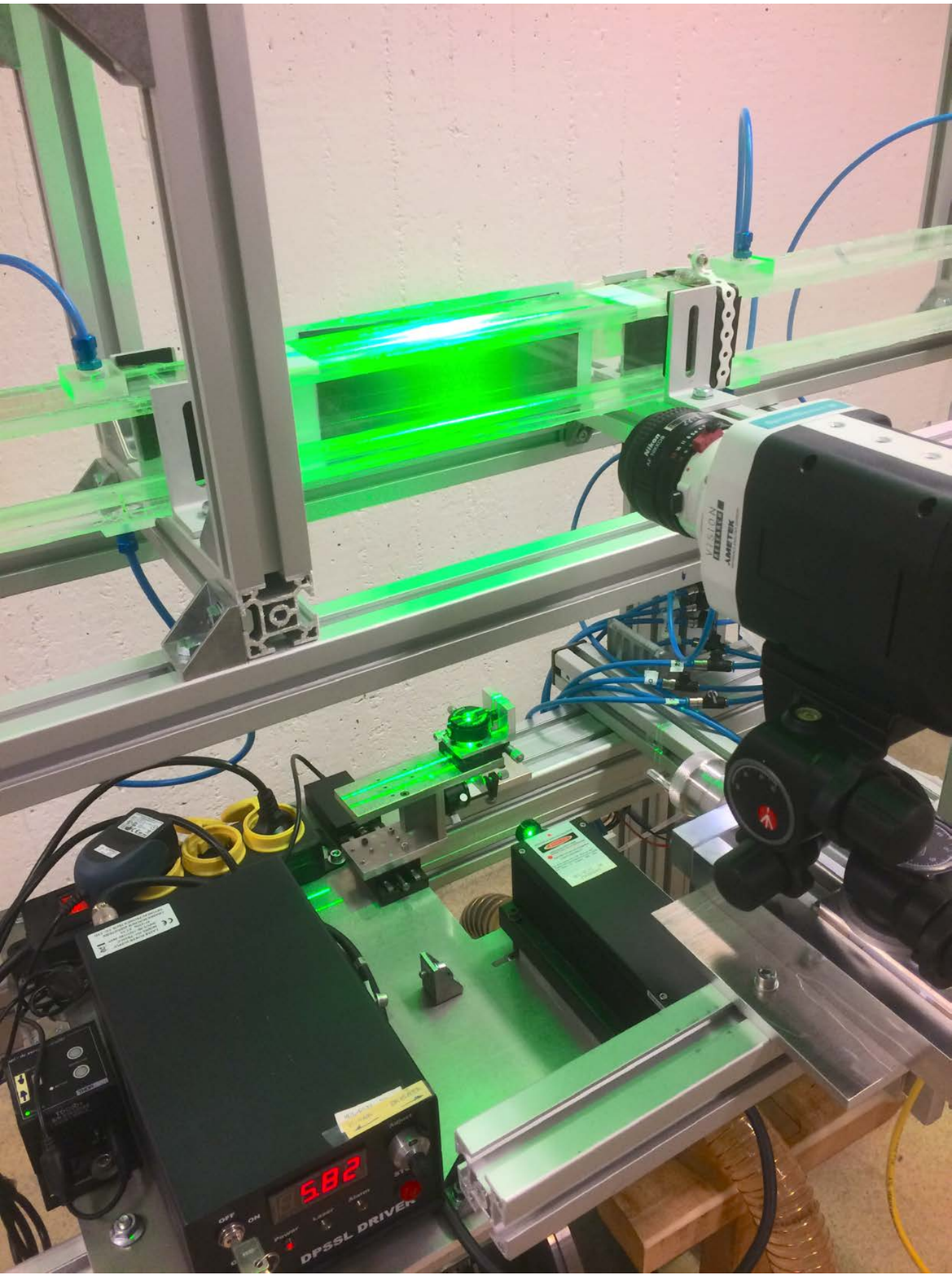}
  \caption{}
  \label{fig:PIV set-up}
\end{subfigure}
\caption{(a) Schematic of the flow-loop (b) Photo of the section where PIV is performed.}
\label{fig:Set-up}
\end{figure}

The experiments were performed in a transparent Plexiglas square duct that is 50 mm x 50 mm in cross section and 5 m in length. Figure \ref{fig:Set-up schematic} shows a schematic of the flow loop. The fluid is recirculated through a conical tank that is open to the atmosphere, where the particle-fluid mixture can be introduced. A tripping tape is lined on the inner walls of the Plexiglas duct at the inlet to trigger turbulence. The temperature of the solution is maintained at nearly 20$^\circ$C by means of an external heat-exchanger in the tank. A gentle disc pump (Discflo Corporations, CA, USA) has been chosen to minimize mechanical breakage of the particles. It can pump rather large particles at reasonably high volume concentration without pulsations. Additional details about the set-up can be found in \cite{zade2018experimental}.

\begin{figure}
\centering
  \includegraphics[width=0.7\linewidth]{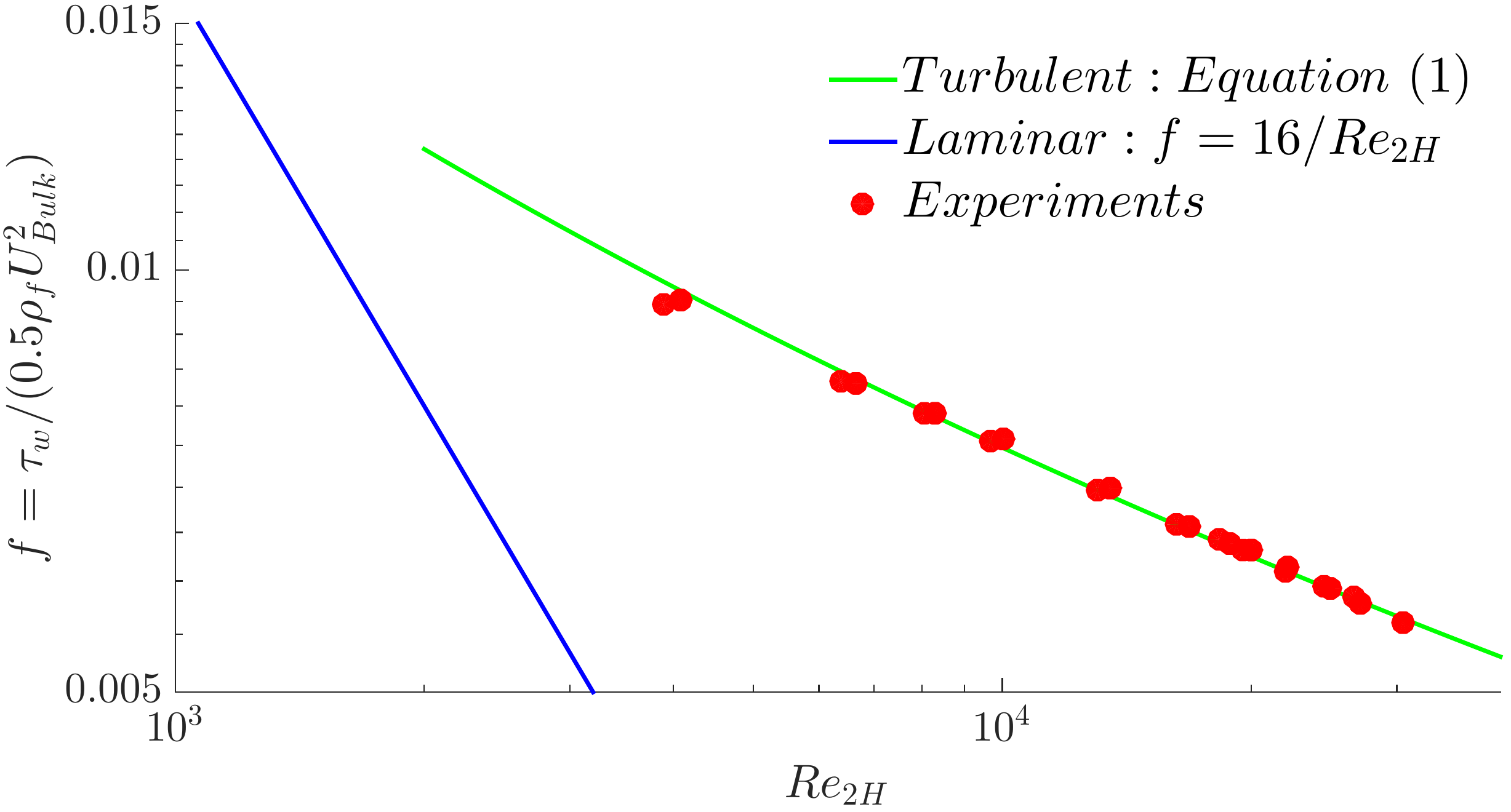}
\caption{Comparison between experiments, DNS simulations and empirical correlation for the Fanning friction factor $f$ as a function of bulk Reynolds number $Re_{2H}$ for single-phase Newtonian fluid. }
\label{fig:Friction factor plot}
\end{figure}

An electromagnetic flowmeter (Krohne Optiflux 1000 with IFC 300 signal converter, Krohne Messtechnik GmbH, Germany) is used to measure the flow rate. The Reynolds number $Re_{2H}$, used hereafter, is based on the average or bulk velocity $U_{Bulk}$ of the fluid-particle mixture, the viscosity of the fluid ${\eta}$ and full height of the duct $2H$. The pressure drop is measured across a length of $54H$ in a region of the duct that is nearly $140H$ from the inlet (the turbulent flow was seen to be fully developed at this entry length) using a differential pressure transducer (0 - 1 kPa, Model: FKC11, Fuji Electric France, S.A.S.). Figure \ref{fig:Friction factor plot} shows a reasonable agreement between the friction-factor $f={\tau_w}/{(\rho_fU_{Bulk}^2/2)}$ for the single phase Newtonian fluid flow measured in our square duct and the empirical correlation given in \cite{duan2012pressure}, 
\begin{equation}
  {f}={\Big(3.6\log_{10}(\frac{6.115}{Re_{\sqrt{A}}})\Big)}^{-2}.
  \label{eqn:Friction factor}
\end{equation}
Here, $\tau_w = (dP/dx)(H/2)$ is the wall shear stress measured from the streamwise pressure gradient $dP/dx$ and $\rho_f$ is the density of the fluid. The Reynolds number $Re_{\sqrt{A}}$ in equation \ref{eqn:Friction factor} is based on the characteristic length given by the square root of the cross section area $A = 2H\times2H$. The friction velocity, used later, is given by $u_{\tau} = \sqrt{\tau_w/\rho_f}$. Data acquisition from the camera, flow meter and pressure transducer is performed using a National Instruments NI-6215 DAQ card using Labview\textsuperscript{TM} software.

\subsection{Particle properties}

The finite-size particles are commercially procured super-absorbent (polyacrylamide based) hydrogel. Once mixed with water and left submerged for nearly 24 hours at room temperature, they grow to an equilibrium size of 5 $\pm$ 0.8 mm. To enhance the contrast of the particles in the PIV images, a small amount of Rhodamine is added to the water in which the particles expand. The particle size is determined by a digital imaging system as well as from the PIV images. The fact that a Gaussian like particle size distribution, with small variance, has small effect on the flow statistics has been shown in \cite{FORNARI201854}. 

The density of the particles is determined by measuring the terminal settling velocity of a single particle with a known diameter, gently dropped in a long liquid settling column. The relation for drag force $F$ on a settling particle in \cite{crowe2011multiphase},
\begin{equation}
  \frac{F}{\rho_{f}U_{T}^2A}=\frac{12}{Re_{p}}(1+0.15Re_{p}^{0.687})
  \label{eqn:Particle density}
\end{equation} applicable in the transitional regime: 1 $ < Re_p < $ 750, is used to relate the particle diameter $d_p$ and terminal velocity $U_T$ to the unknown particle density $\rho_{p}$. Here, $A$ is the projected area of the particle and $Re_p$ is the particle Reynolds number given by $\frac{\rho_{p}U_{T}d_{p}}{\mu_f}$ where, ${\mu_f}$ is the dynamic viscosity of the liquid. The measurements were repeated multiple times at a room temperature of around 20$^\circ$C and yielded a particle to fluid density ratio $\rho_p/\rho_f$ = 1.0035$\pm$0.0003.

For experiments in NF, if pure water is used as the suspending fluid, the particle density ratio is nearly equal to one, suggesting neutrally-buoyant particles i.e.\ the influence of gravity on the particle motion is negligible. However, at $Re_{2H}\approx$ 11000, some sedimentation effects were visible in water. This can be quantified by calculating the Rouse number $Ro={U_T}/{\kappa u_\tau}$ \citep{rouse1937}, which is used to estimate whether the particles move as a bed load ($Ro\geq$ 2.5) or in full suspension ($Ro\leq$ 1.2) \citep{fredsoe1992mechanics}. For water, at a $Re_{2H}\approx$ 11000, the $Ro\approx$ 2.4, which leads to settling. Hence, to ensure that particles are in full suspension at the given $Re_{2H}$, it is necessary to increase the viscosity, so that the $U_{Bulk}$ proportionately increases, while maintaining the density ratio nearly equal to one. This is accomplished by adding 3.6\% by mass of low molecular weight (MW = 8000) Polyethylene Glycol (PEG) (Carbowax\textsuperscript{TM}, Fischer Scientific) to water which resulted in a transparent Newtonian solution with a viscosity that is 2.2 times the viscosity of water and $Ro\approx$ 0.8 which corresponds to particles in full suspension. 

\subsection{Polymer rheology}

\begin{figure}
\centering
\includegraphics[height=0.50\linewidth]{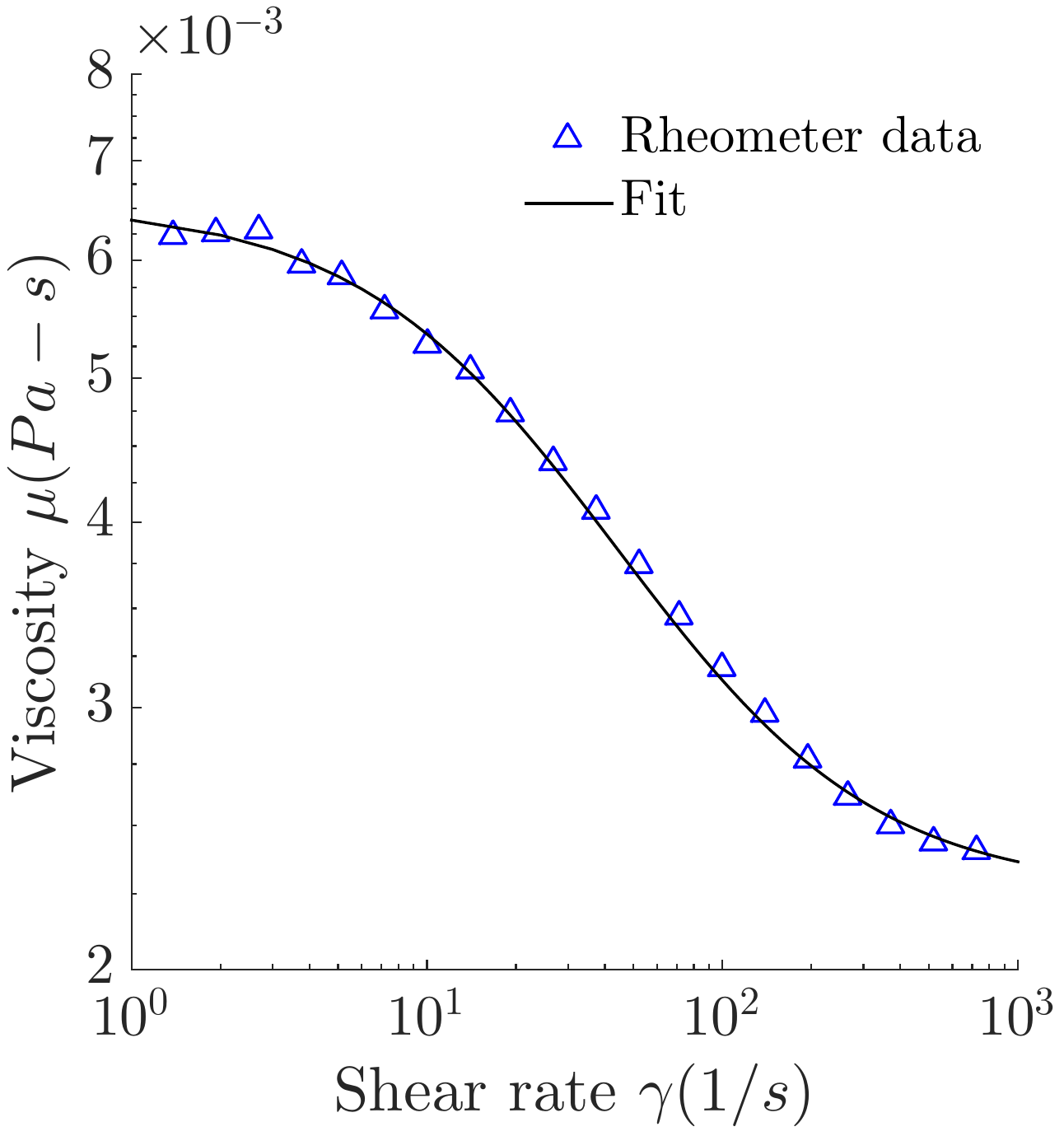}
\caption{Variation of viscosity with shear rate for the viscoelastic fluid.}
\label{fig:sr_vs_mu}  
\end{figure}

The VEF solution was prepared by adding 250 ppm of a high molecular weight polyacrylamide based anionic polymer (FLOPAM AN934SH, SNF) to water. The density of the solution is practically the same as the solvent (water) under such dilute concentrations. The solution was prepared by dissolving the polymer in powder form to water, followed by successive dilution and mixing to ensure its homogeneity. Such a solution with the desired polymer concentration was stored in a large reservoir from which it was pumped to the tank at the beginning of every new experiment. The shear viscosity of the polymer solution was measured using a rheometer (Kinexus pro+, Malvern Panalytical). The variation of dynamic viscosity with shear rate is shown in Figure \ref{fig:sr_vs_mu} and shear thinning behavior can be clearly observed. It is known that only shear-thinning by itself produce no drag-reduction and hence, viscoelasticity is important \citep{metzner1964turbulent}. \sz{From the study of \citet{owolabi2017turbulent}, who used the same type and concentration of polymer additive as in this study, it can be said that the present level of drag reduction (at the same mass flow rate) corresponds to a $Wi$ between 1--2.} For this non-Newtonian fluid, the Reynolds number $Re_{2H} = U_{Bulk}2H/\eta_{w}$, where $\eta_{w}$ is the near-wall viscosity which corresponds to the average experimental shear stress (or equivalently shear rate) obtained from the pressure drop measurements in the fully developed region of the duct.

\subsection{Velocity measurement technique}

\begin{figure}
\centering
\begin{subfigure}{.3\textwidth}
  \centering
  \includegraphics[height=0.95\linewidth]{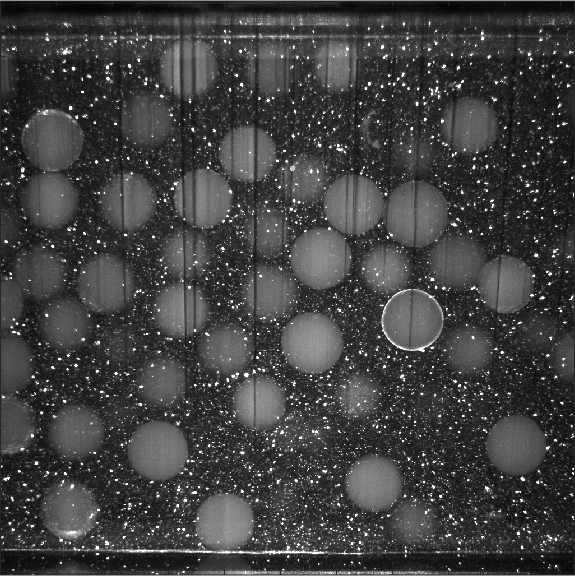}
  \caption{}
  \label{fig:PIV image}
\end{subfigure}%
\begin{subfigure}{.3\textwidth}
  \centering
  \includegraphics[height=0.95\linewidth]{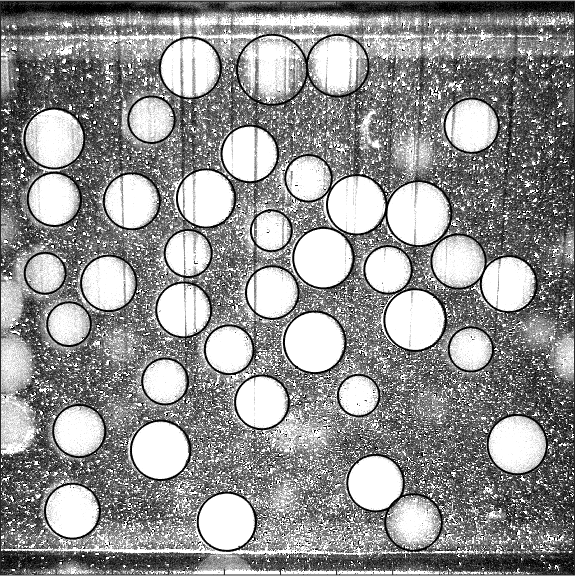}
  \caption{}
  \label{fig:PTV image}
\end{subfigure}
\begin{subfigure}{.3\textwidth}
  \centering
  \includegraphics[height=0.95\linewidth]{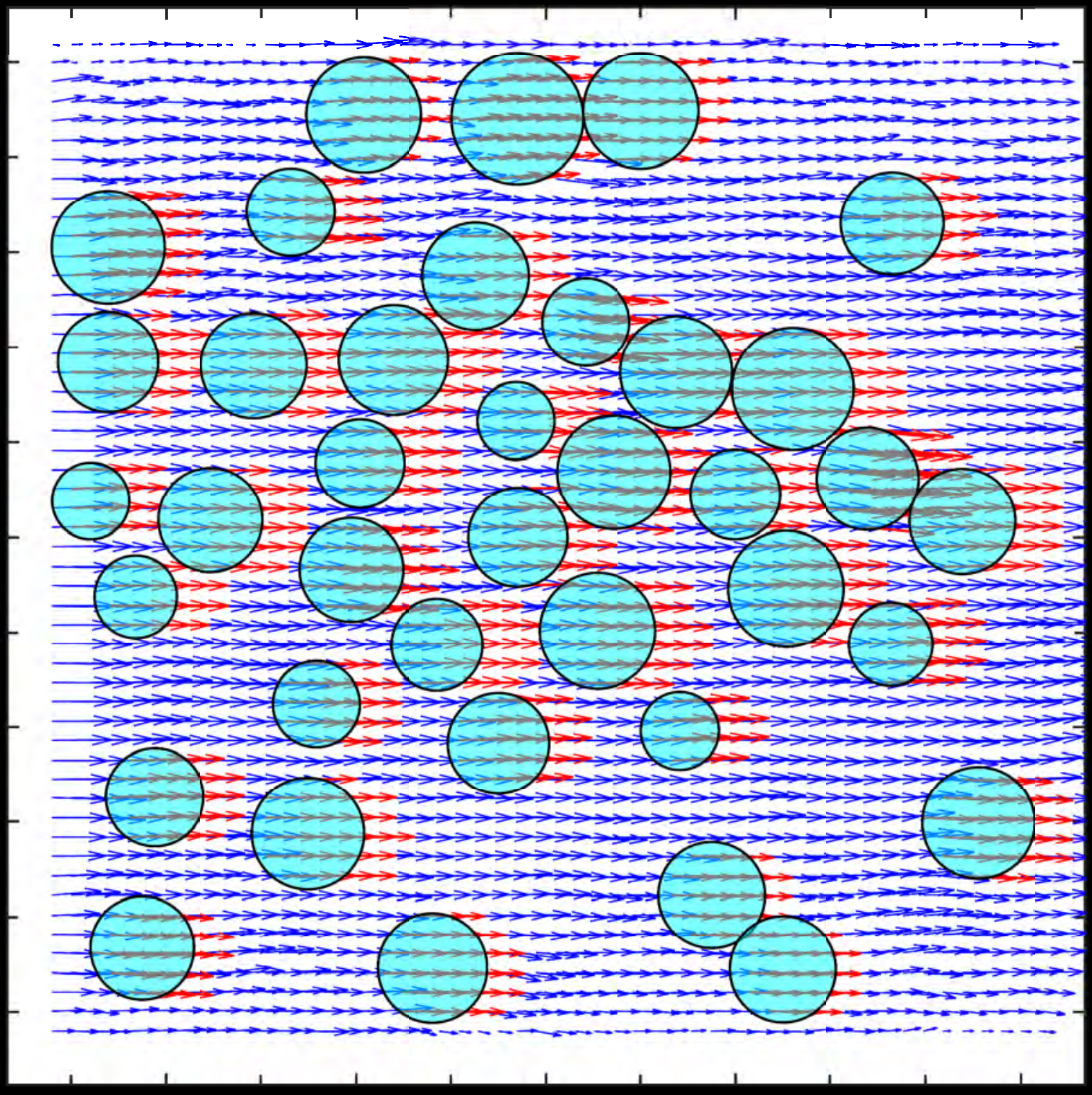}
  \caption{}
  \label{fig:PIV+PTV}
\end{subfigure}
\caption{(a) Image for PIV analysis, (b) image for particle detection and PTV analysis and (c) combined fluid PIV - particle PTV velocity vectors for $\phi$ = 20\%.}
\label{fig:PIV+PTV image and velocity vectors}
\end{figure}

The coordinate system used in this study is indicated in figure \ref{fig:Set-up schematic} with $x$ the streamwise, $y$ the wall-normal and $z$ the spanwise directions. The velocity field is measured using 2D Particle Image Velocimetry (2D-PIV) in the plane of the wall-bisector: $z/H$ = 0. Thus, the two lateral walls are situated at $z/H$ = -1 and 1 respectively. These measurements are performed at a streamwise distance of $x/H\approx$ 150 from the entrance of the duct. A continuous wave laser (wavelength = 532 nm, power = 2 W) and a high-speed camera (Phantom Miro 120, Vision Research, NJ, USA) are used to capture successive image pairs. The thickness of the laser light-sheet is 1 mm. Figure \ref{fig:PIV set-up} shows a photo of the PIV set-up. For imaging the full height of the duct, a resolution of approximately 60 mm/1024 pixels is chosen. The frame rate (acquisition frequency) is selected such that the maximum pixel displacement does not exceed a quarter of the size of the final interrogation window IW \cite{raffel2013particle}. Images are processed using an in-house, three-step, FFT-based, cross-correlation algorithm \cite{kawata2014velocity}. The final size of the IW is 32 $\times$ 32 pixel. The degree of overlap can be estimated from the fact that the corresponding final resolution is 1 mm x 1 mm per IW. Each experiment has been repeated at least 2 times and 500 image pairs, each separated by more than 4 flow turn-over time $T = 2H/U_{Bulk}$ so as to ensure statistically independent samples, have been observed to be sufficient for statistically converged results both for single phase and particle-laden flows.

Figure \ref{fig:PIV+PTV image and velocity vectors} depicts one image from a typical PIV sequence for particle-laden flow. Raw images captured during the experiment are saved in groups of two different intensity levels. The first group of images (an example being figure \ref{fig:PIV image}) is used for regular PIV processing according to the algorithm mentioned above. The second group of images (cf. figure \ref{fig:PTV image}) are contrast-enhanced, e.g. they are sharpened and their intensity adjusted, and used for detecting the finite-size particles only using a circular Hough transform \cite{yuen1990comparative}. From the detected particles in image A and B of the PIV sequence, a nearest neighbor approach is used to determine their translational motion. Particles that are detected only in one image of the pair are, thus, eliminated by the PTV algorithm. For the Eulerian PIV velocity field, we define a mask, which assumes the value 1 if the point lies inside the particle and 0, if it lies outside. The fluid phase velocity is thus determined on a fixed mesh. The particle velocity is determined using PTV at its center, which is assigned to the grid points inside the particle (mask equal to 1). The velocity field of the particle-phase is, now, available at the same grid points as that of the fluid and the ensemble averaging, reported later, are phase averaged statistics. Figure \ref{fig:PIV+PTV} shows the combined fluid (PIV) and particle (PTV) velocity field. A point to note is that, using the above PTV approach, we could measure the translational velocity of the particle but particle rotation could not be measured.

\section{Results}

\subsection{Single phase flow}

\begin{figure}
\centering
\begin{subfigure}{.49\textwidth}
  \centering
  \includegraphics[height=1\linewidth]{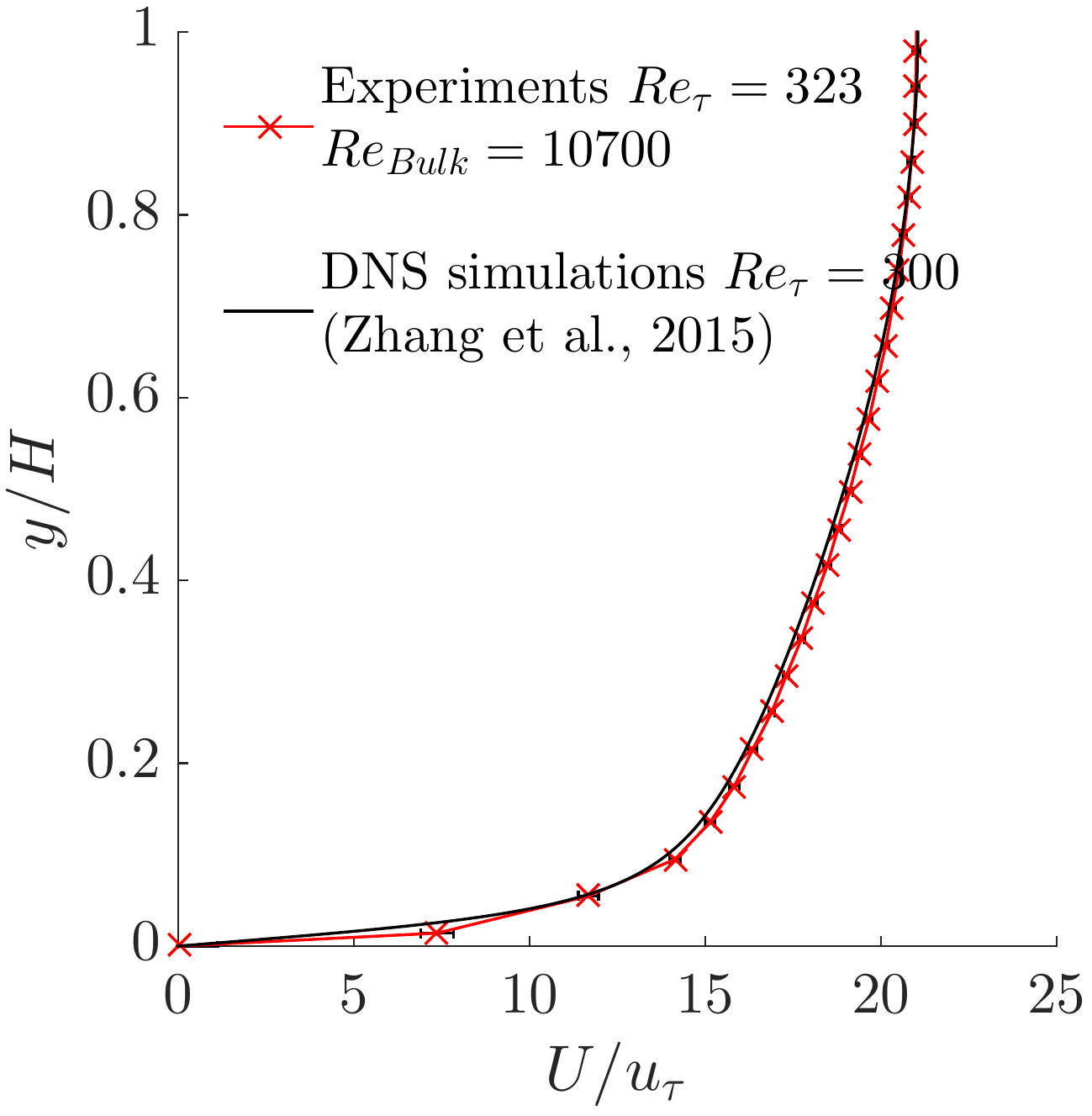}
  \caption{}
  \label{fig:U_sp_N}
\end{subfigure}%
\begin{subfigure}{.49\textwidth}
  \centering
  \includegraphics[height=1\linewidth]{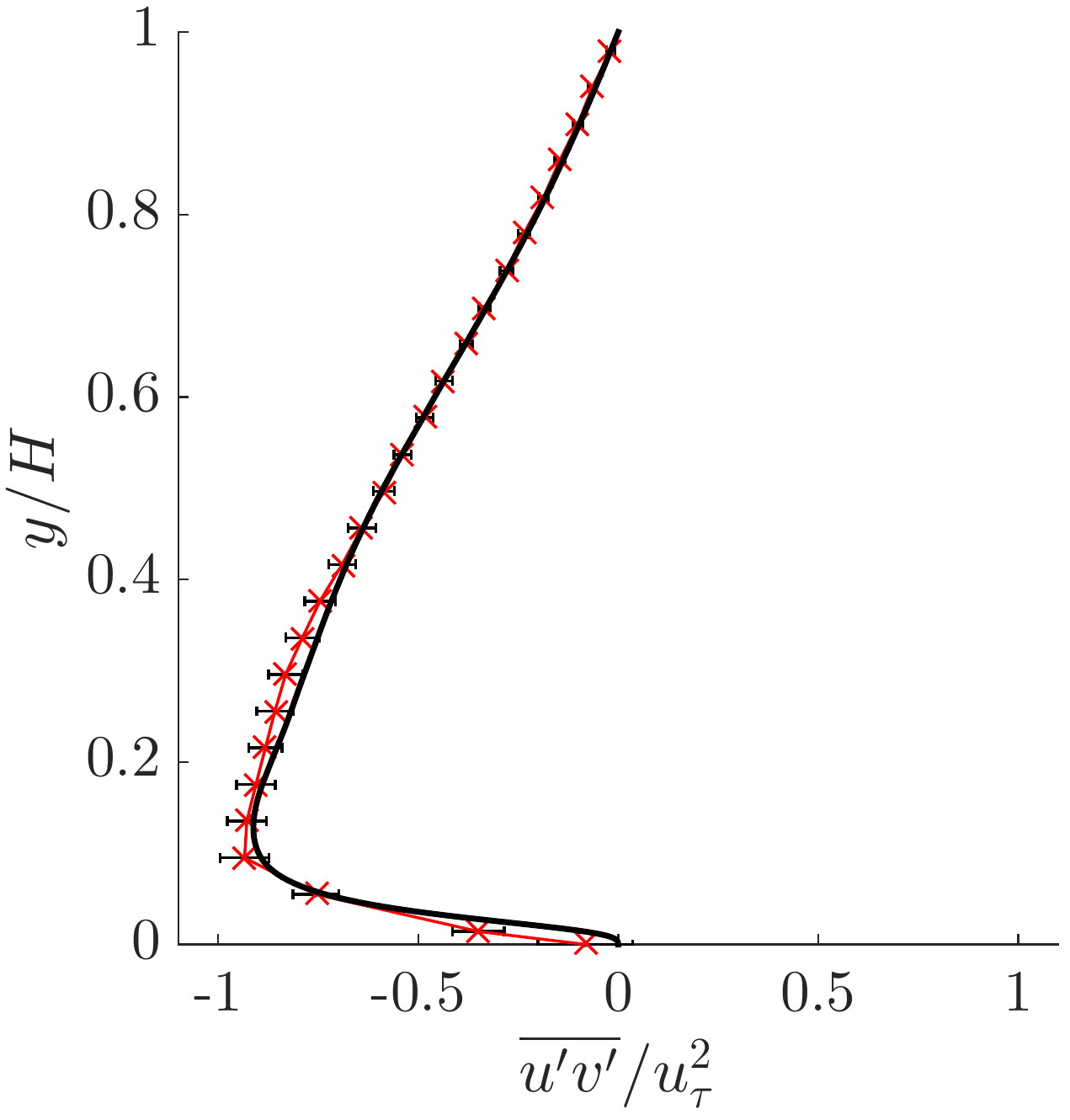}
  \caption{}
  \label{fig:uv_sp_N}
\end{subfigure}
\caption{Single phase Newtonian fluid (NF): (a) mean stream-wise velocity profiles and (b) Reynolds shear stress in the plane of the wall-bisector compared with the DNS simulations of Zhang et al., 2015 at a slightly lower $Re_{\tau}$.} 
\label{fig:U_mean_and_uv_sp_Newt.}
\end{figure}

Figure \ref{fig:U_mean_and_uv_sp_Newt.} shows the mean streamwise velocity profile and the Reynolds shear stress in the plane of the wall-bisector for single phase flow of NF at $Re_{2H}$ = 10700$\pm$100 which corresponds to $Re_{\tau}$ = 323$\pm$6. Only the bottom half is shown due to symmetry. Error bars with a width of two standard deviations is also plotted for the experimental data. Comparison with DNS simulations of \citet{zhang2015direct} at a slightly lower $Re_{\tau}$ = 300 shows reasonable agreement both in terms of mean velocity and correlation of fluctuating velocity components. 

\begin{figure}
\centering
\begin{subfigure}{.49\textwidth}
  \centering
  \includegraphics[height=1\linewidth]{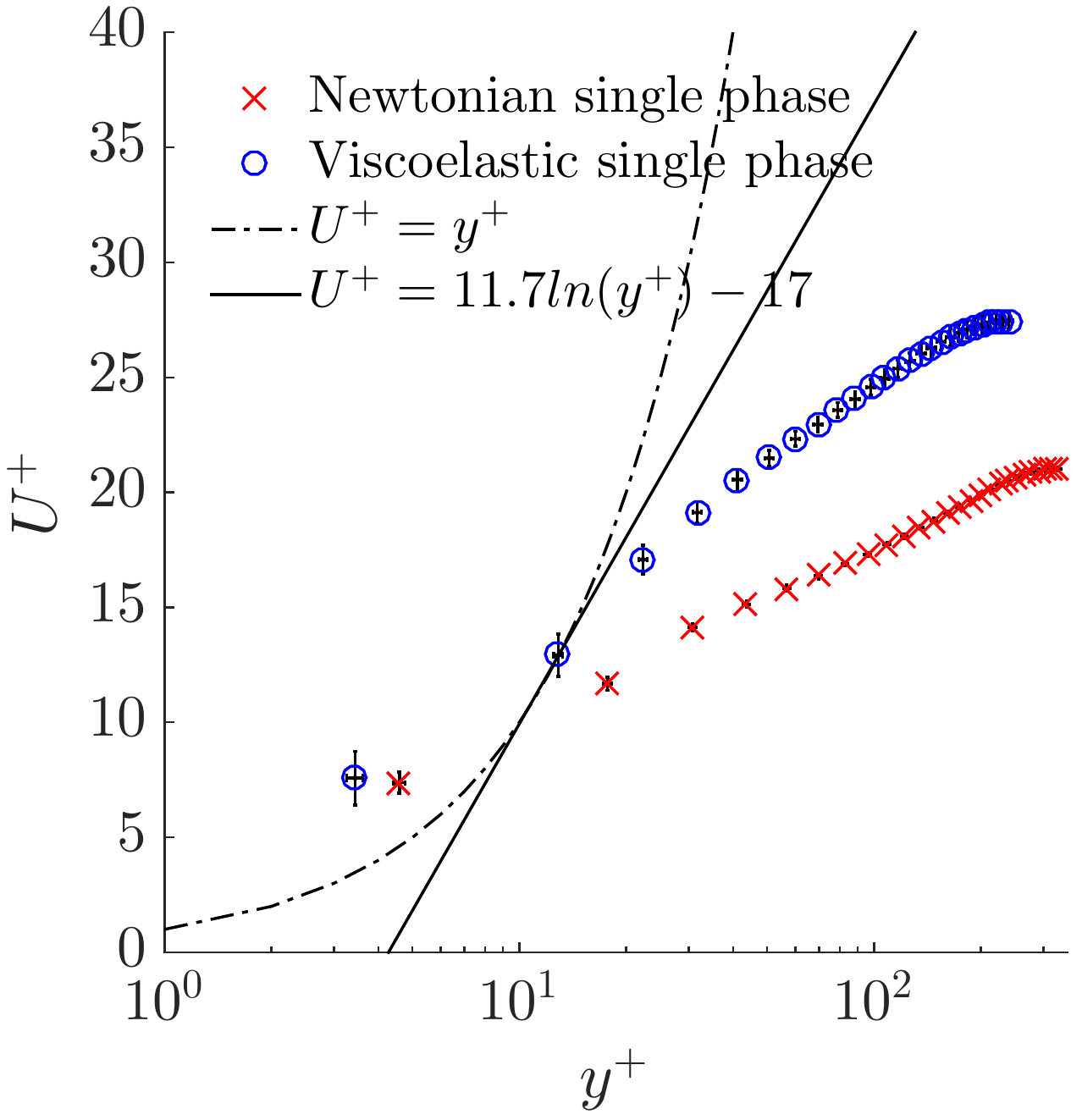}
  \caption{}
  \label{fig:Uplus_sp_N}
\end{subfigure}
\begin{subfigure}{.49\textwidth}
  \centering
  \includegraphics[height=1\linewidth]{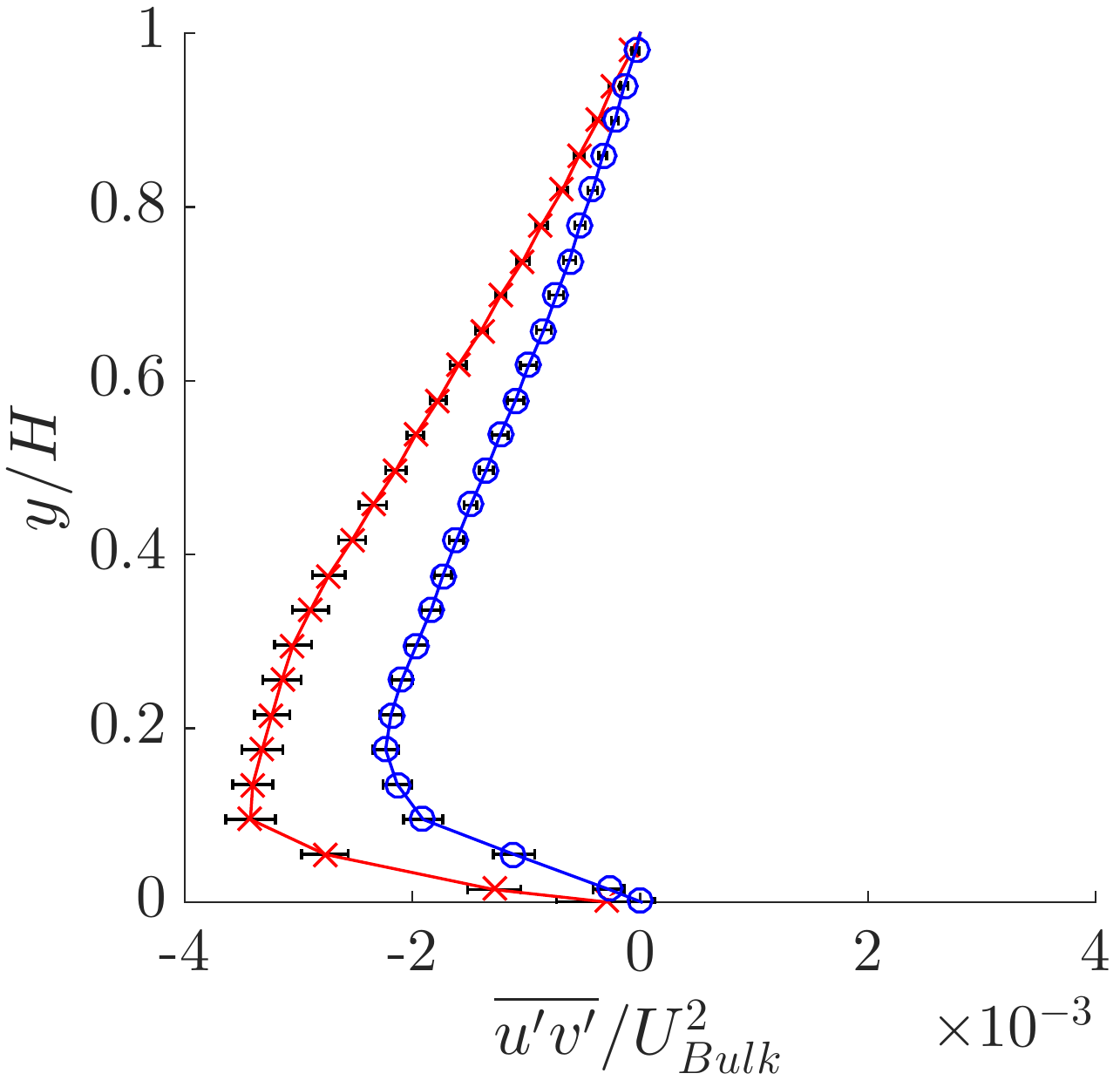}
  \caption{}
  \label{fig:uv_sp_N}
\end{subfigure}
\begin{subfigure}{.49\textwidth}
  \centering
  \includegraphics[height=1\linewidth]{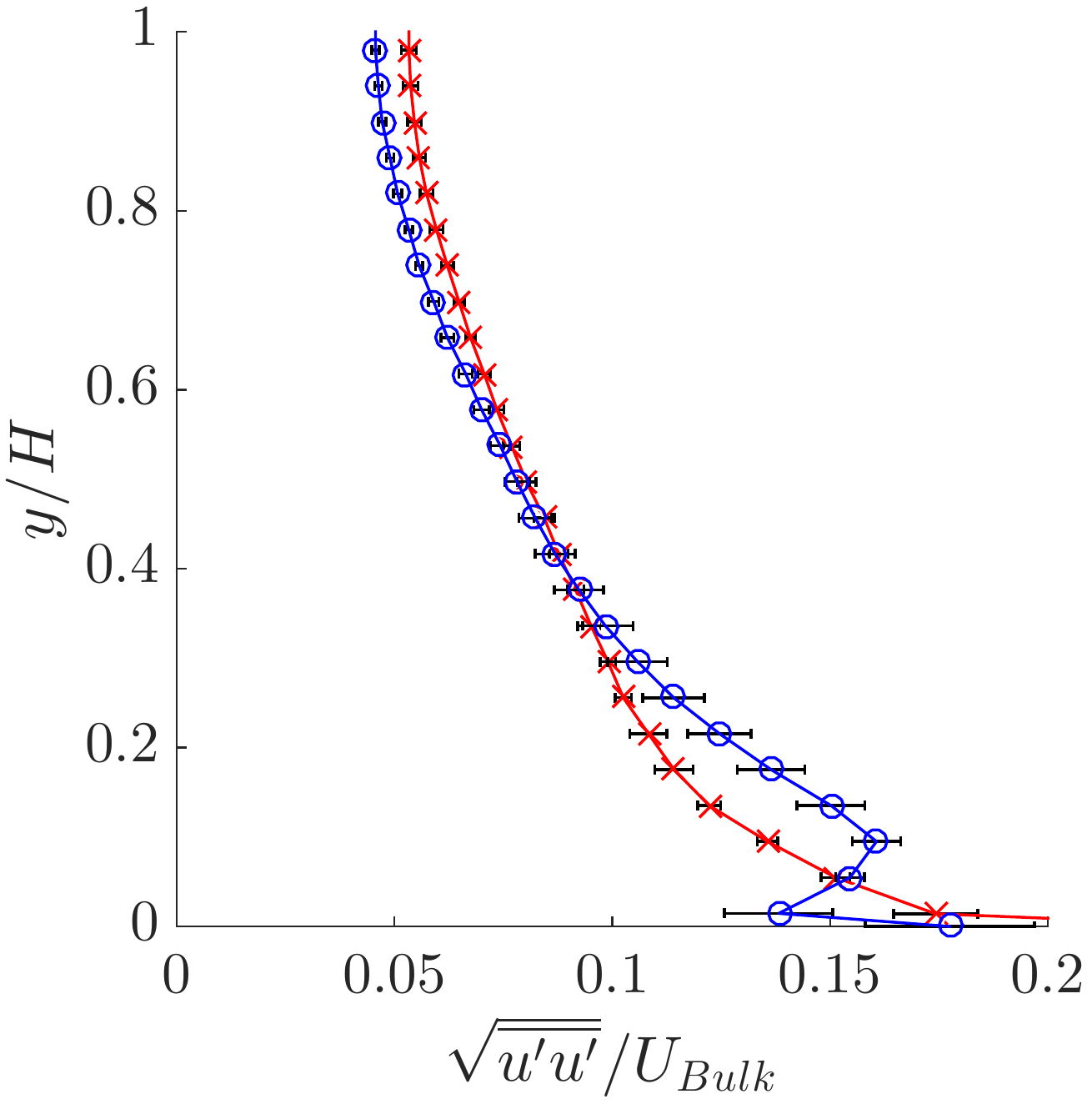}
  \caption{}
  \label{fig:uu_sp_N}
\end{subfigure}
\begin{subfigure}{.49\textwidth}
  \centering
  \includegraphics[height=1\linewidth]{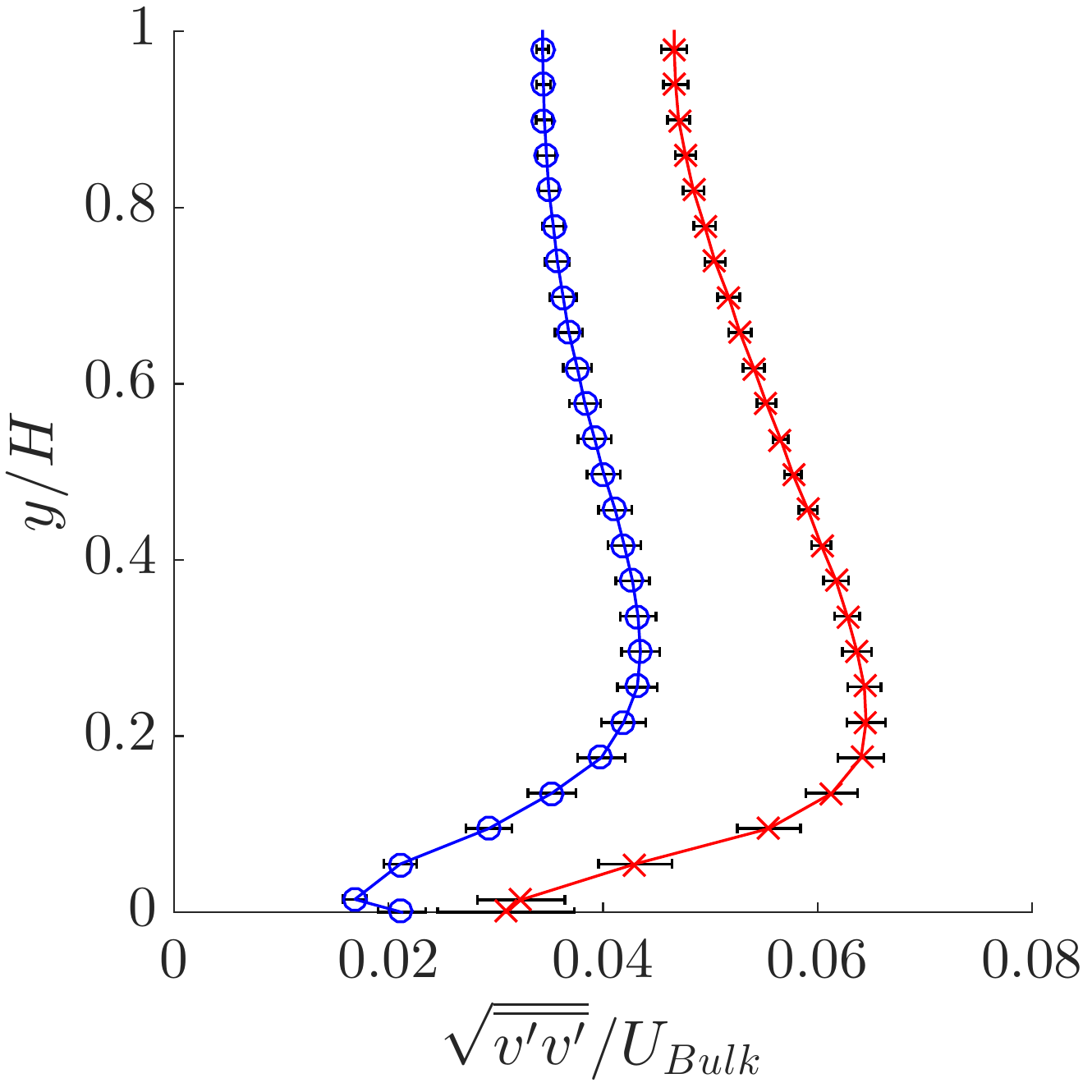}
  \caption{}
  \label{fig:vv_sp_N}
\end{subfigure}
\caption{Comparing single phase NF and VEF flow: (a) mean stream-wise velocity profiles, (b) Reynolds shear stress and fluid velocity fluctuations in the (c) stream-wise and (d) wall-normal directions in the plane of the wall-bisector.} 
\label{fig:Velocity_Newtonian_vs_DR}
\end{figure}

\begin{figure}
\centering
\begin{subfigure}{.49\textwidth}
  \centering
  \includegraphics[height=1\linewidth]{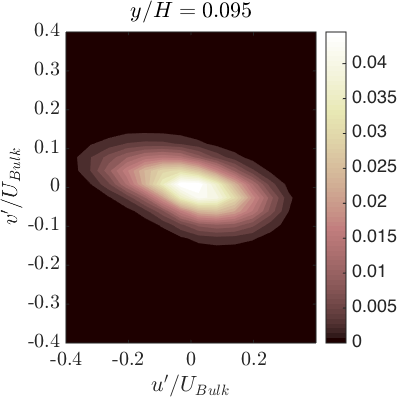}
  \caption{}
  \label{fig:JPDF_Newt}
\end{subfigure}%
\begin{subfigure}{.49\textwidth}
  \centering
  \includegraphics[height=1\linewidth]{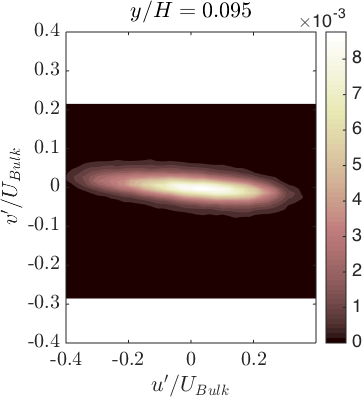}
  \caption{}
  \label{fig:JPDF_Poly}
\end{subfigure}
\caption{Joint probability distribution function of (a) single phase NF, (b) VEF at $y/H$ = 0.095. The joint PDF has been normalized such that the sum of the PDF is equal to 1.} 
\label{fig:JPDF}
\end{figure}

Addition of polymer additives introduces elasticity in the fluid phase and modifies the flow and overall drag and velocity statistics. For nearly the same $Re_{2H}$ = 10200$\pm$100, drag for the single phase VEF flow is 43$\pm$2\% lower than for the single phase NF. Drag change is calculated as the relative change in friction factor $f$ as compared to friction factor of single phase Newtonian flow $f_{Sp, NF}$, at the same $Re_{2H}$ and is given by
\begin{equation}
  Relative\ drag\ modification=\left(100\times\frac{f_{Sp, NF} - f}{f_{Sp, NF}}\right)_{Re_{2H} = constant}
  \label{eqn:Drag change}
\end{equation}
The drag variation for the particle-laden cases is estimated in the same way. 
Figure \ref{fig:Velocity_Newtonian_vs_DR} shows the turbulent velocity statistics of single phase NF and VEF. The mean streamwise velocity, scaled in inner units, is shown in figure \ref{fig:Uplus_sp_N}. The average $u_{\tau}$ is measured from the pressure drop, and the wall-normal distance $y$ is scaled using $\eta_w/u_{\tau}$, where $\eta_w$ is the viscosity corresponding to the average shear stress at the wall, as mentioned before. Virk's ultimate profile corresponding to the Maximum Drag Reduction (MDR) asymptote \citep{virk1975drag} is also shown. The velocity profile is shifted upwards with a slightly higher slope than the Newtonian case in the log-region. Our resolution close to the wall is not sufficient to collapse data on $U^+ = y^+$ in the viscous sub-layer $y^+ \leq 5$. The mean streamwise velocity if scaled in bulk units (shown later in figure \ref{fig:U_fluid_NN}) does not show any major differences compared to the Newtonian case, except in the near-wall region where the NF is slightly faster due to the higher level of turbulent mixing driven by the corresponding higher Reynolds shear stress, see figure \ref{fig:uv_sp_N}, which shows that the streamwise and wall-normal fluctuations are less correlated for turbulent VEF. The streamwise velocity fluctuations scaled by bulk quantities or equivalently turbulent intensity increase (cf figure \ref{fig:uu_sp_N}) below $y/H$ = 0.4 and marginally reduce above $y/H$ = 0.4. The location of the peak in the turbulence intensity, which correlates to the location of maximum turbulence production, shifts away from the wall for the VEF. The corresponding peak for NF is at $y^+\approx$ 15 \citep{zhang2015direct} or equivalently $y/H \approx$ 0.05, which is almost the first measurement point of our PIV and hence, cannot be fully captured at the resolution that we use. This results in the false visual impression that the streamwise turbulence intensity is maximum at the wall for NF, where it should have been zero. The wall-normal velocity fluctuations decrease (cf figure \ref{fig:vv_sp_N}) in VEF causing an increased anisotropy as has been almost universally observed \citep{gyr2013drag} for drag-reducing turbulent flows. Increase in the streamwise fluctuations with polymer additives is a characteristic of the low drag reduction regime as proposed in \citet{warholic1999influence} whereas, for high drag reduction, streamwise fluctuations reduce and the location of the peak is shifted further away from the wall. Thus, the turbulent fluctuations are predominantly streamwise, as also seen by the joint Probability Distribution Function (PDF) of the fluctuating streamwise and wall-normal velocity in figure \ref{fig:JPDF} (at a near wall location $y/H$ = 0.095). The higher streamwise alignment of the major axis of the joint PDF for VEF in figure \ref{fig:JPDF_Poly} compared to NF in figure \ref{fig:JPDF_Newt} clearly indicates that the turbulent fluctuations are preferentially streamwise. In other words there is less momentum transferred towards the walls, which is equivalent to drag reduction.

\subsection{Particles in Newtonian flow}

\subsubsection{Drag modulation}

\begin{figure}
\centering
\includegraphics[height=0.50\linewidth]{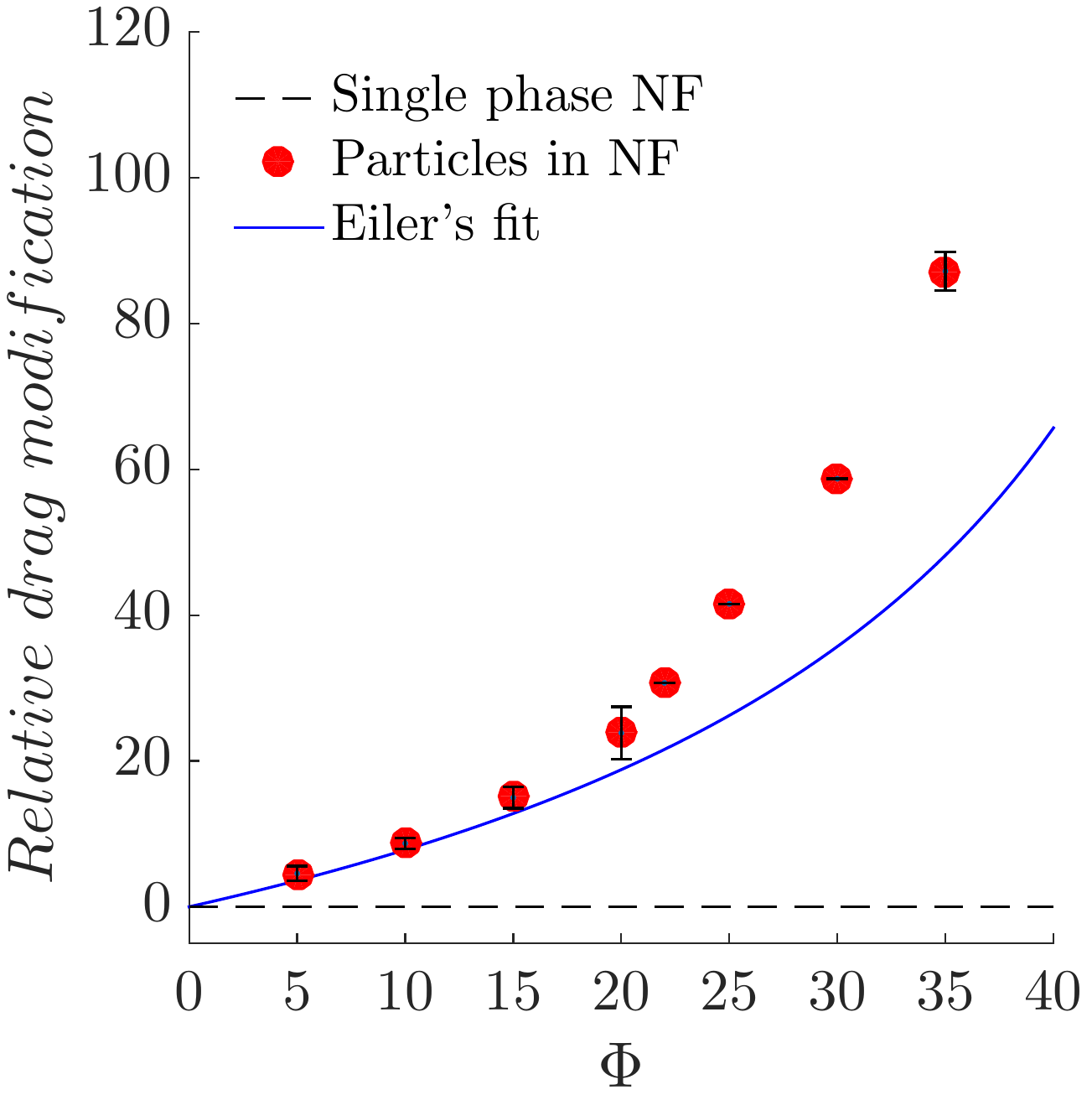}
\caption{Relative drag modification versus particle volume fraction $\phi$ in Newtonian fluid (NF) at $Re_{2H}$ = 10700$\pm$150.}
\label{fig:Phi_Drag_Newtonian}  
\end{figure}

Figure \ref{fig:Phi_Drag_Newtonian} shows the change in drag (cf equation \ref{eqn:Drag change}) for varying volume fraction $\phi$ of particles in a NF at $Re_{2H}$ = 10700. The drag increases with increasing $\phi$. The solid line shows the change in drag that can be predicted by an effective suspension viscosity, obtained from the Eilers fit \citep{stickel2005fluid}; this empirical formula relates the effective viscosity to the nominal volume fraction $\phi$ in the limit of vanishing inertia,
\begin{equation}
  \frac{\eta_e}{\eta} = \Big(1 + \frac{\frac{5}{4}\phi}{1-\frac{\phi}{0.65}}\Big)^2.
  \label{eqn:Eilers fit}
\end{equation}
The effective viscosity $\eta_e$ is used to compute an effective Reynolds number $Re_e=U_{Bulk}2H/\eta_e$, in turn used to find the effective friction factor from equation \ref{eqn:Friction factor}. This is then compared with a single phase NF having the viscosity of the suspending solution $\eta$. This simple approach predicts an increase in the drag with the particle concentration, although of different magnitude than that observed experimentally. The prediction is closer to the experimental values for low $\phi \leq$ 10\% and diverges substantially with increasing $\phi$. \citet{abbas2017pipe} studied a laminar flow of concentrated non-colloidal particles ($\phi$ = 70\%) and used the notion of effective viscosity, based on the local particle concentration to explain the observed pressure drop. Also, recently, \citet{bakhuis2018finite} found a net drag increase for an increase in $\phi$ in a different geometry, Taylor-Couette flow, at very high $Re$. However, the increase was much smaller than predicted by the increase in effective viscosity due to the particles. From the above examples, it is clear that the pressure drop cannot be correctly estimated using an effective viscosity formulation corresponding to the nominal $\phi$ and perhaps, spatial variation in $\phi$ needs to be considered in the spirit of \citet{costa2016universal} who proposed scaling laws for the mean velocity profile of the
suspension flow. These authors also calculated an equation able to predict the increase in drag as a function of the particle size and volume fractions in a channel flow. Their theory assumes that the flow domain can be split into two regions: (i) a region close to the wall where the difference between the mean velocity of the two phases is  substantial and (ii) a region away from the wall, where the mean flow is well represented by the continuum limit of a Newtonian fluid with an effective viscosity.

\subsubsection{Velocity statistics}

\begin{figure}
\centering
\begin{subfigure}{.49\textwidth}
  \centering
  \includegraphics[height=1\linewidth]{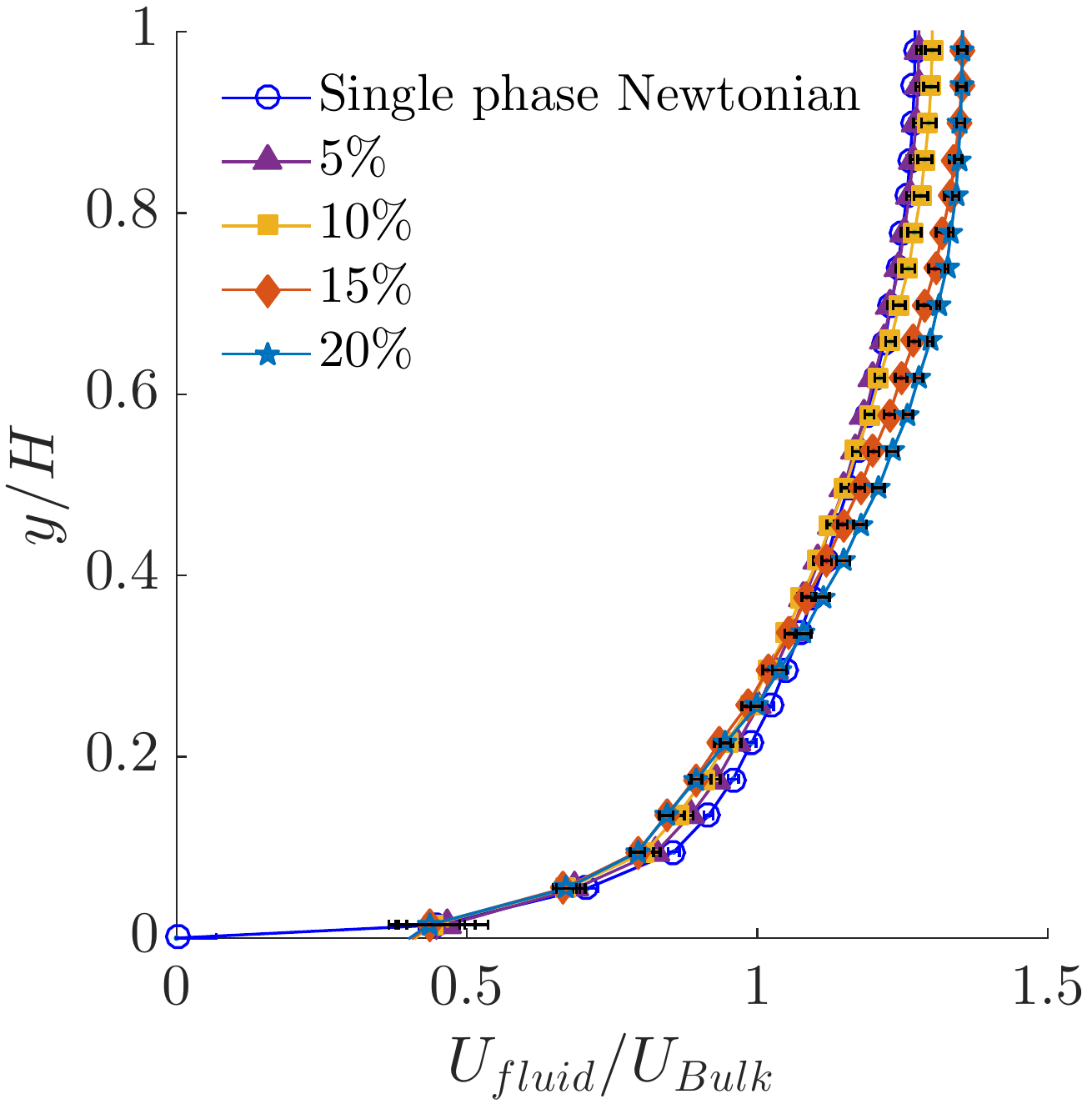}
  \caption{}
  \label{fig:U_fluid}
\end{subfigure}%
\begin{subfigure}{.49\textwidth}
  \centering
  \includegraphics[height=1\linewidth]{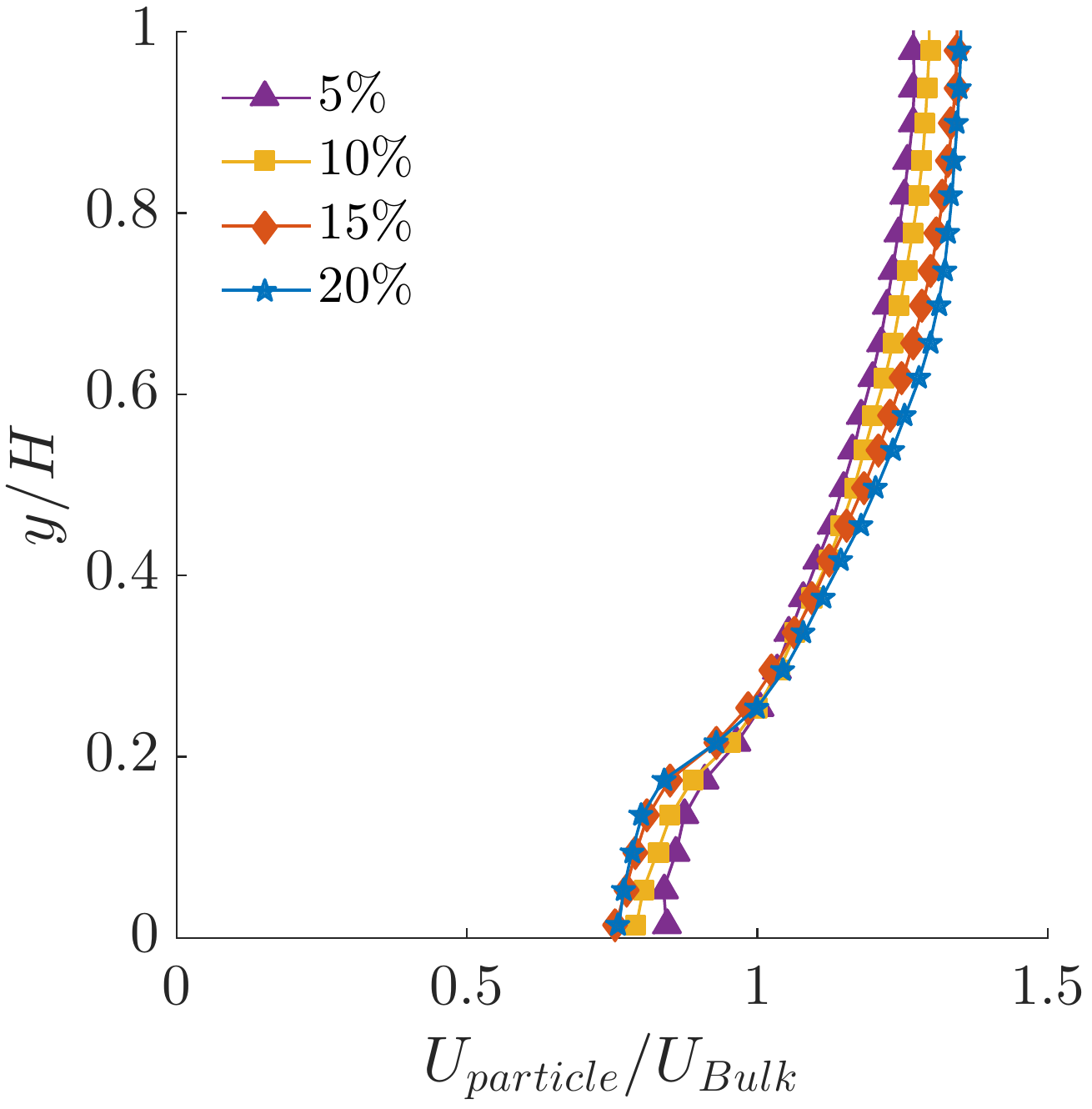}
  \caption{}
  \label{fig:U_particle}
\end{subfigure}
\begin{subfigure}{.49\textwidth}
  \centering
  \includegraphics[height=1\linewidth]{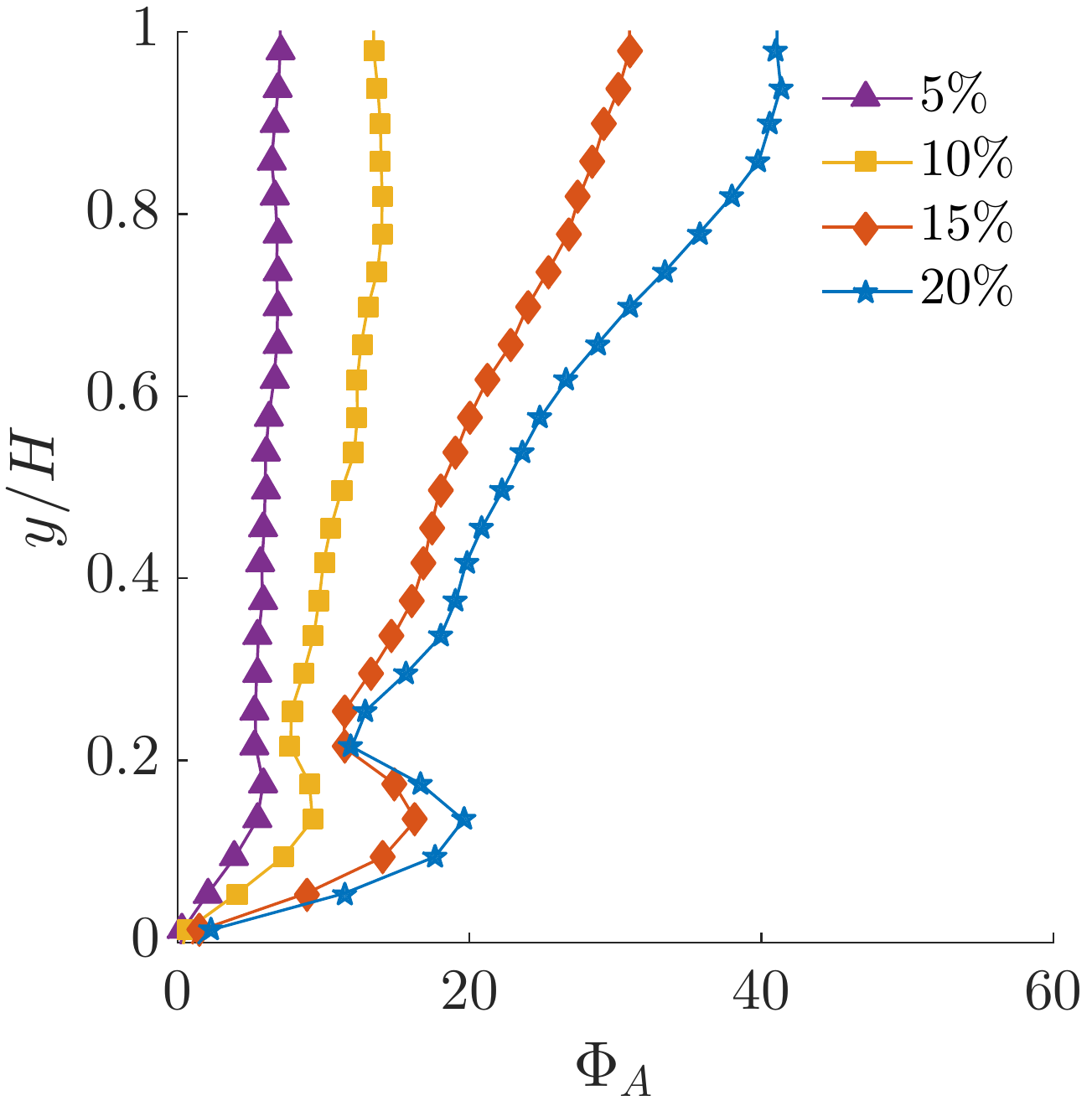}
  \caption{}
  \label{fig:Phi}
\end{subfigure}
\caption{Particle-laden Newtonian fluid: Mean stream-wise velocity profiles for (a) fluid phase and (b) particle phase. The particle (area) concentration profile is shown in (c)}
\label{fig:U_mean_and_Phi_Newt.}
\end{figure}

Figure \ref{fig:U_fluid} shows the mean fluid streamwise velocity for different $\phi$ in NF. The ratio of maximum velocity $U_{Max}$, at the center of the duct $y/H$ = 1, to $U_{Bulk}$ increases as $\phi$ increases. \sz{It is known that in the single phase turbulent regime, with decreasing $Re_{2H}$ the ratio $U_{Max}/U_{Bulk}$ increases due to reduced cross-stream mixing. Thus, increasing $\phi$ modifies the streamwise velocity profile in a way similar to the reduction of $Re_{2H}$.} From the velocity profile, it appears that the mean streamwise velocity gradient increases with increasing $\phi$ which suggests that the contribution of fluid viscous stress $\mu_f dU/dy$ to the overall stress increases with increasing $\phi$, at least in the plane of the wall-bisector.

The particle velocity profile in figure \ref{fig:U_particle} exhibits a large apparent slip velocity in the near-wall region. This value in the near-wall region is most likely over-estimated because we could not measure the rotational velocity of the particle and the PTV measurement assumes that the entire particle is translating with the velocity of the centroid. In the flow, however, the particle also rotates, more in regions of higher shear rate, and hence the slip velocity will be lower in the near-wall region than that shown in figure \ref{fig:U_particle}. Away from the walls, the particle rotation is lower due to lower shear rate and the estimate of the mean velocity is closer to the true value. The particle mean streamwise motion closely follows the fluid mean streamwise motion away from the near-wall region. 

The particle concentration distribution profile in figure \ref{fig:Phi} shows characteristic maxima at the center as well the presence of a particle-rich layer near the wall as previously observed \citep{costa2017finite, fornari2016effect}. At $\phi$ = 5\%, the particles are nearly uniformly distributed with a very weak indication of local maxima. With increasing $\phi$, particles tend to migrate preferentially towards the core. This could be due to inertial shear-induced migration as explained in \citet{fornari2016effect} or/and due to imbalance in the normal stresses in the wall-normal direction as seen in \citet{lashgari2016channel}. The local undulations in the concentration profile, especially for the highest $\phi$ = 20\%, occurs due to particle layering, are visible due to the rather large size of our particles ($2H/d_p$ = 10).

\begin{figure}
\centering
\begin{subfigure}{.49\textwidth}
  \centering
  \includegraphics[height=1\linewidth]{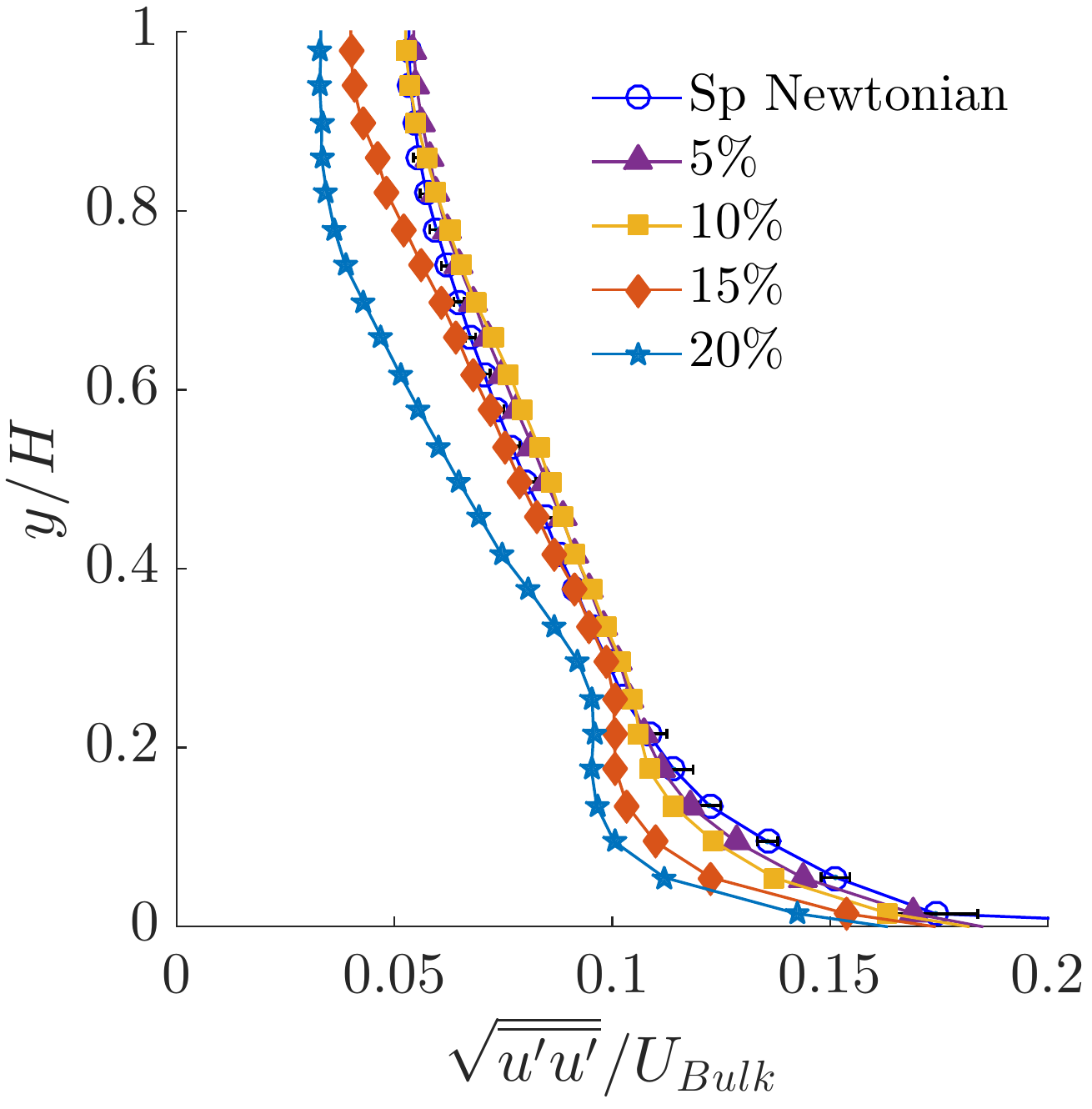}
  \caption{}
  \label{fig:urms}
\end{subfigure}%
\begin{subfigure}{.49\textwidth}
  \centering
  \includegraphics[height=1\linewidth]{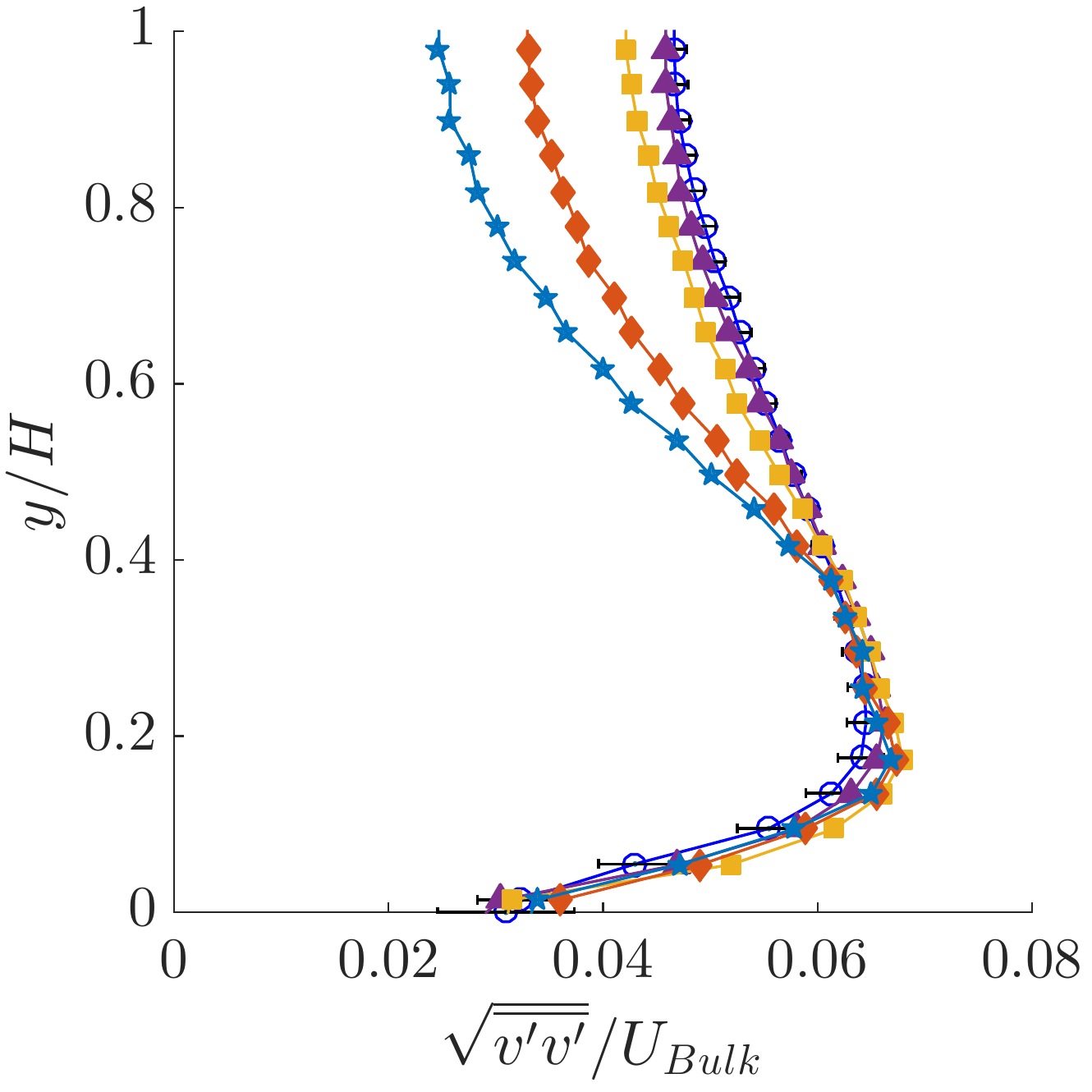}
  \caption{}
  \label{fig:vrms}
\end{subfigure}
\begin{subfigure}{.49\textwidth}
  \centering
  \includegraphics[height=1\linewidth]{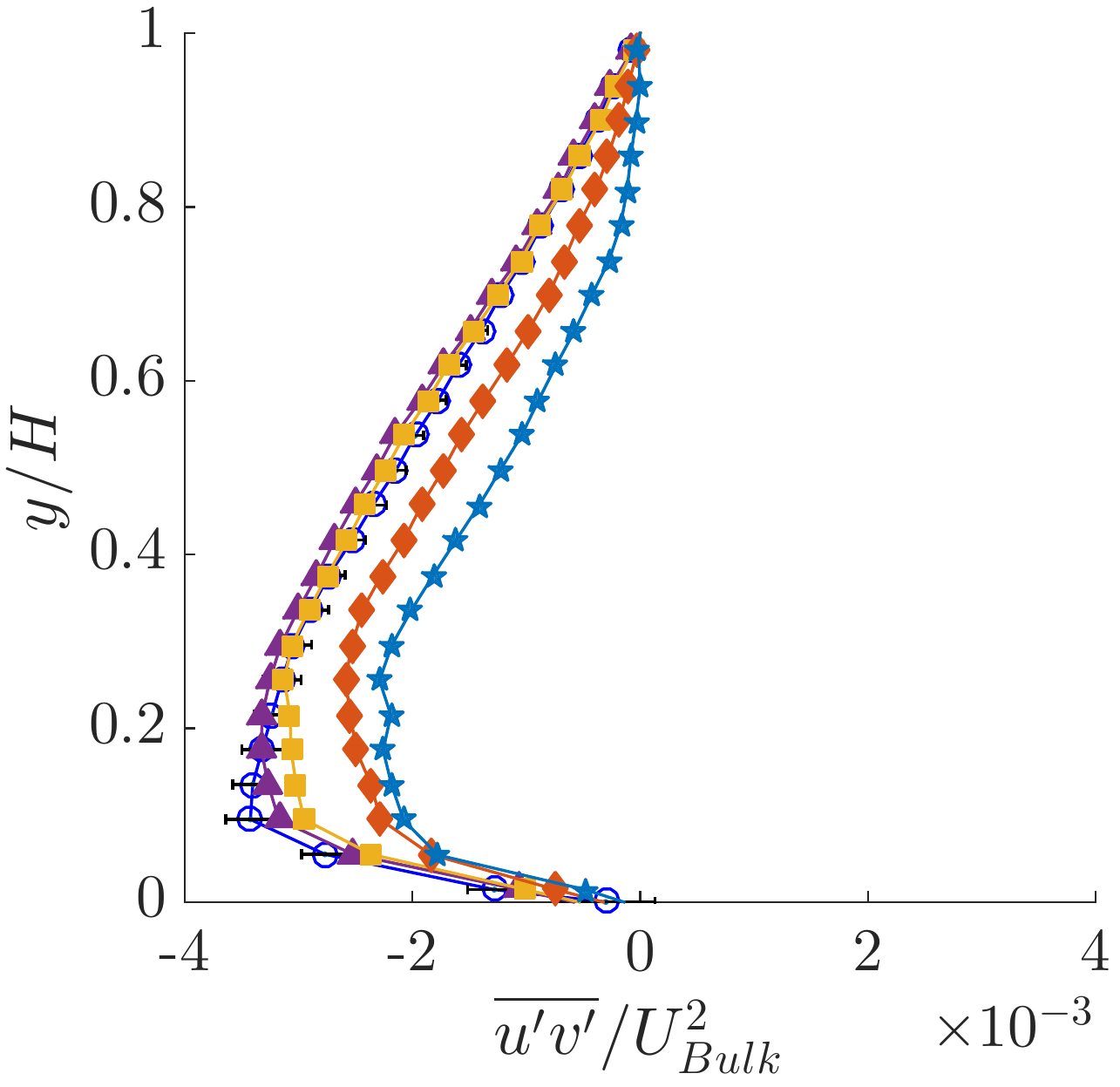}
  \caption{}
  \label{fig:uv}
\end{subfigure}
\begin{subfigure}{.49\textwidth}
  \centering
  \includegraphics[height=1\linewidth]{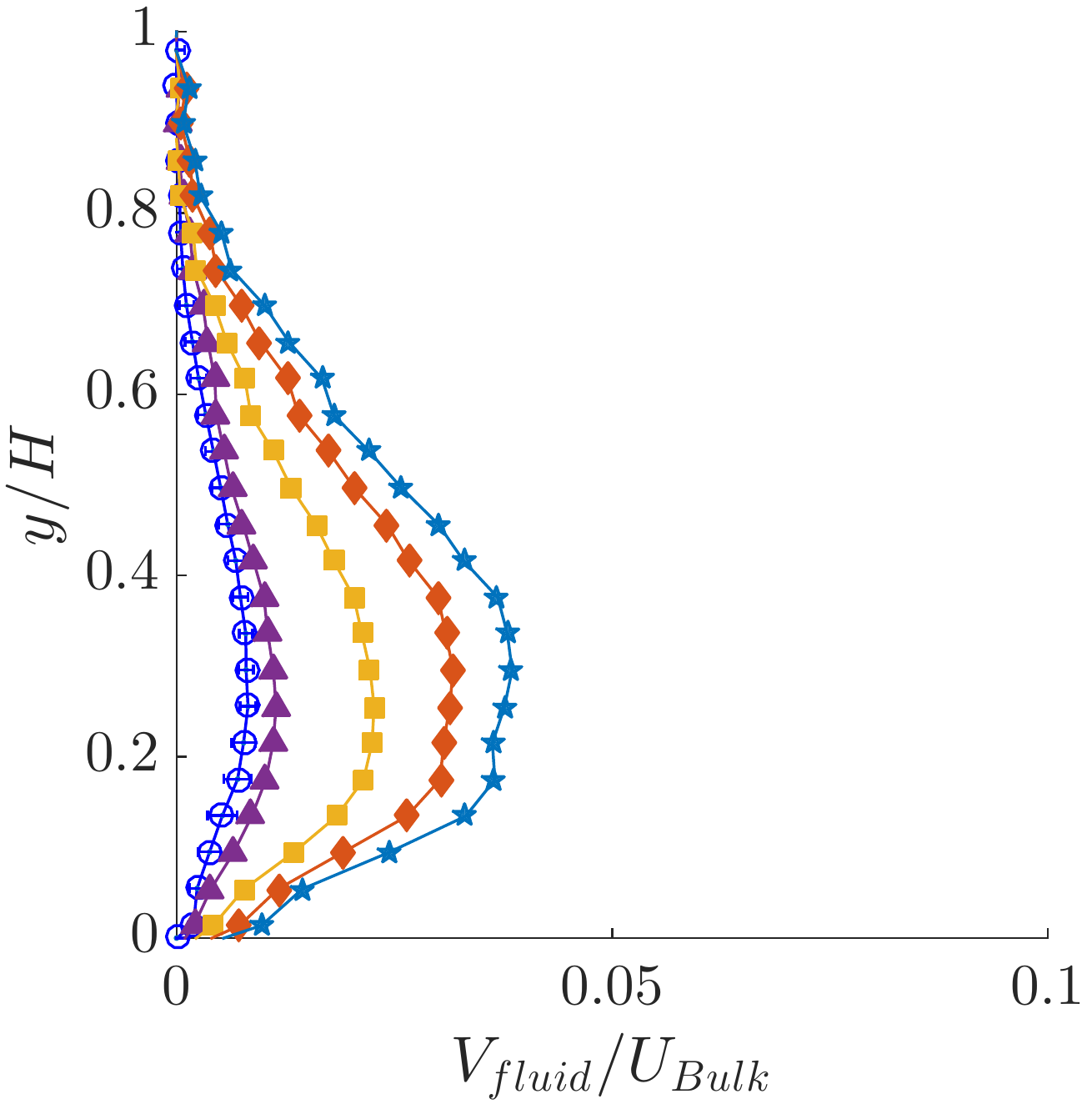}
  \caption{}
  \label{fig:V}
\end{subfigure}
\caption{Particle-laden Newtonian fluid: Fluid velocity fluctuations in the (a) stream-wise direction and (b) wall-normal direction and (c) Reynolds shear stress. Panel (d) depicts the mean wall-normal velocity. Sp stands for Single phase.}
\label{fig:fluctuating_velocity}
\end{figure}

The fluctuations in the fluid velocity at different $\phi$ are shown is shown in figure \ref{fig:fluctuating_velocity}. The streamwise turbulence intensity is shown in figure \ref{fig:urms}. The inadequate resolution, as mentioned before, leads to a spurious high value of the turbulence intensity at the wall. Nevertheless, the values at higher $y/H$ clearly indicate a reduction in the streamwise velocity fluctuations in a statistically significant sense. For $\phi$ = 5 and 10\%, compared to the single phase NF case, the streamwise velocity fluctuations marginally increase above the wall-normal location $y/H \approx$ 0.2 while they decrease below $y/H \approx$ 0.2. Such a behaviour was also observed, albeit more pronounced, in \cite{fornari2017suspensions} at $\phi$ = 5 and 10\%. With further increase in $\phi$, the streamwise velocity fluctuations reduce for all wall-normal locations in our study. Contrarily, in \cite{fornari2017suspensions} the turbulence intensity at $\phi$ = 20\% increases in an intermediate region in between the wall region and the core. We believe that the above differences are consistent with the larger turbulence attenuation caused by our larger particles ($D/d_p$ = 10) compared to the smaller particles ($D/d_p$ = 18) in the simulations. Also, the smaller Reynolds number $Re_{2H}$ = 5600 in their simulations (refer to \cite{fornari2017suspensions}) compared to $Re_{2H}$ = 11000 in the present study may lead to differences.

The wall-normal fluctuations (cf figure \ref{fig:vrms}) are lower in magnitude as compared to their streamwise counterpart for single phase NF flow. With addition of particles, these wall-normal fluctuations further reduce, monotonically with $\phi$. The reduction near the core is substantial and, noticeably, the peak value remains nearly constant for all $\phi$. \sz{This damping of the wall-normal velocity fluctuations in the core region further indicates reduction of turbulence by particles in that region.}

The primary Reynolds shear stress, shown in figure \ref{fig:uv}, reduces with increasing $\phi$ in the near wall region ($y/H\leq$0.2), meaning that the correlation between fluid streamwise and wall-normal velocity reduces with increasing particle concentration. This can be explained by the disruption of coherent near-wall structures, responsible for the peak in single-phase flow, by the particle-rich near-wall layer. It can be noted that for low $\phi$ = 5 and 10\%, the primary Reynolds shear stress marginally increase above $y/H \approx$ 0.2. As also mentioned in \cite{bakhuis2018finite}, one can speculate that particles up to a certain $\phi$ introduce wakes i.e.\ coherent flow structures in the mean flow leading to an increase in the correlation between fluid streamwise and wall-normal velocity. However, with increasing $\phi$, the wakes from particles will interact with one another resulting in reduced correlation. At the highest $\phi$ = 20\%, the lower Reynolds shear stress (figure \ref{fig:uv}) and lower mean streamwise velocity gradient (figure \ref{fig:U_fluid}) may cause lower production of the streamwise velocity fluctuations as seen in figure \ref{fig:urms}.

Turbulent square duct flow exhibits secondary flow of Prandtl's second kind, driven by gradients in the turbulence-stresses, especially the secondary Reynolds shear stress and the second normal stress difference \citep{gavrilakis1992numerical}. These secondary motions, in the form of four pairs of counter-rotating vortices located at the duct corners act to transfer fluid momentum from the center of the duct to its corners, thereby causing a bulging of the streamwise mean velocity contours toward the corners. The strength of this secondary flow is weak, generally between 1 to 4\% of the bulk velocity in most straight ducts with non-circular cross-section, and hence prone to larger measurement uncertainties. Figure \ref{fig:V} shows the mean wall-normal velocity profile, originating due to the secondary flow, in the plane of the wall-bisector $z/H$ = 0. With increasing $\phi$, the magnitude of the secondary flow progressively increases, at least in the plane of the wall-bisector. \citet{fornari2017suspensions} associated the increased secondary flow with the larger gradient in the second normal stress difference as well as fluid-particle momentum exchange for $\phi$ = 5 and 10\% for their smaller particles. However, at $\phi$ = 20\%, they observed a reduction in the mean secondary flow, contrary to our observation. Thus, it appears that the modulation of secondary flow depends on the particle size and it continues to increase with $\phi$, up to a certain $\phi$ which is a function of particle size.

\subsection{Particles in drag reducing flow}

In the following sections, we report the data for particles in a VEF at nearly the same $Re_{2H}$ as in the NF cases described previously.

\subsubsection{Change in drag}

\begin{figure}
\centering
\includegraphics[height=0.50\linewidth]{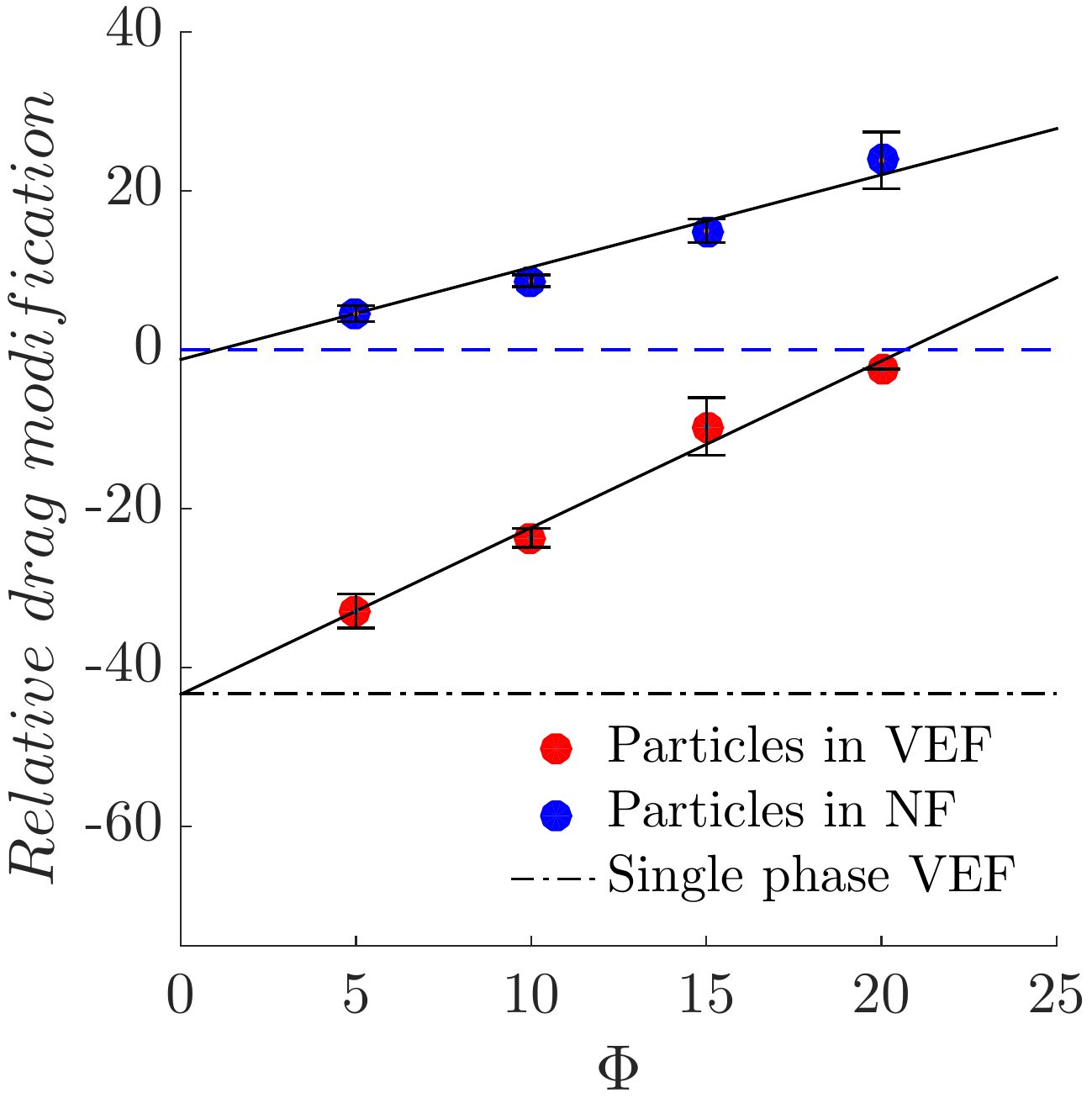}
\caption{Relative drag modification versus particle volume fraction for both Newtonian (NF) and viscoelastic fluid (VEF) at nearly the same $Re_{2H}$. For NF, $Re_{2H}$ = 10700$\pm$150 and for VEF, $Re_{2H}$ = 11100$\pm$130.}
\label{fig:Phi_Drag_NN}  
\end{figure}

\begin{table}[b]
\centering
\begin{tabular}{l|lllll}
Cases                                                                                                        & $\phi$  = 0\%   & 5\%               & 10\%                  & 15\%               & 20\% \\
\hline\hline
VEF, $2H/d_p$ = 10, $Re_{2H, \eta_S}$ = 29800$\pm$300       & -26                   & -14                 & -2                        & +15                   & +25 \\  
\hline
NF, $2H/d_p$ = 9, $Re_{2H, \eta_S}$ = 27195 (from \cite{zade2018experimental})                           & 0                       & -1                   & -0.5                    & --                      & +8    \\ 
% VEF, $2H/d_p$ = 10, $Re_{2H, \eta_S} \approx $ 29800$\pm$300       & -26$\pm$2     & -14$\pm$2   & -2$\pm$2        & +15$\pm$4            & +25 \\   
\end{tabular}
\caption{Relative drag modification versus particle volume fraction, at approximately same flow rate and particle size, for Newtonian (NF) and viscoelastic fluid (VEF).}
\label{tab:f_same_Q}
\end{table}

Similar to figure \ref{fig:Phi_Drag_Newtonian}, figure \ref{fig:Phi_Drag_NN} shows the drag change as a function of $\phi$, now for particles in a VEF. The corresponding drag change for particles in a NF is also shown for comparison. As in NF, the drag increases with increasing $\phi$, and nearly approaches the value corresponding to single phase NF at the highest $\phi$ = 20\%. Figure \ref{fig:Phi_Drag_NN} also shows the linear fit for both types of suspending fluid. Clearly, the rate of drag increase with $\phi$ is higher for suspension in VEF. As mentioned before, \cite{murch2017growth} and \cite{einarsson2018einstein} have already observed reduced particle mobility and shear thickening due to elastic effects, which may hint towards a possible explanation for the higher rate of drag increase.

As mentioned before, the change in drag, as represented in figure \ref{fig:Phi_Drag_Newtonian}, is calculated according to equation \ref{eqn:Drag change}. For fluids with shear-dependent viscosity, the drag reduction due to polymer additives is often expressed in terms of the relative change in the friction factor compared to the Newtonian case at the same flow rate \citep{gyr2013drag, owolabi2017turbulent}, instead of the same $Re_{2H}$. This is equivalent to calculating the drag change at a constant Reynolds number $Re_{2H, \eta_S}$, based on the solvent viscosity $\eta_S$. The $Re_{2H, \eta_S}$ for the VEF, now based on $\eta_S$ is 29800$\pm$300, and compared to single phase NF at the same $Re_{2H, \eta_S}$, the drag is 26\% lower. Table \ref{tab:f_same_Q} shows the relative change in drag caused by addition of particles compared to single phase NF at the same flow rate. It can be seen that the drag for $\phi > $ 10\% in VEF is higher than the single phase NF. It is worth noting that in our previous study \cite{zade2018experimental}, in NF with slightly larger particles ($2H/d_p$ = 9) at a slightly lower $Re_{2H, \eta_S}$ = 27195 (cf table \ref{tab:f_same_Q}), we observed a negligible change in drag at $\phi$ = 5 and 10\% compared to single phase NF. For $\phi$ = 20\%, the drag increase was around 8\%, which is still less than the 25\% increase that we observe in VEF for the present case. This higher level of drag increase for particles in VEF compared to NF is relevant from an engineering perspective as it shows that for increasing particle concentration at constant flow rate, addition of polymer may not result in lower pressure drop (or pumping power), compared to particles in Newtonian fluids.

\subsubsection{Velocity statistics}

\begin{figure}
\centering
\begin{subfigure}{.49\textwidth}
  \centering
  \includegraphics[height=1\linewidth]{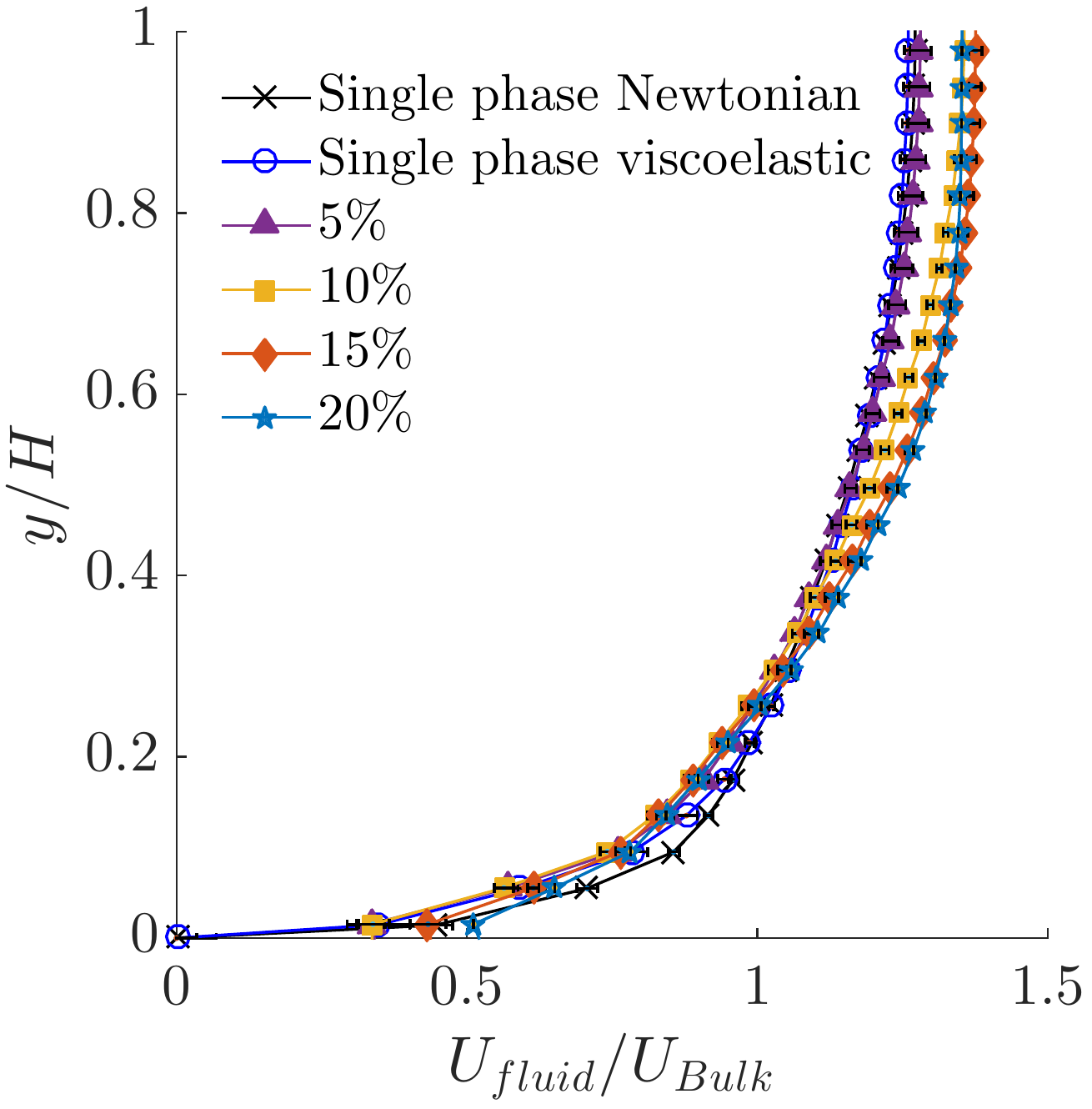}
  \caption{}
  \label{fig:U_fluid_NN}
\end{subfigure}%
\begin{subfigure}{.49\textwidth}
  \centering
  \includegraphics[height=1\linewidth]{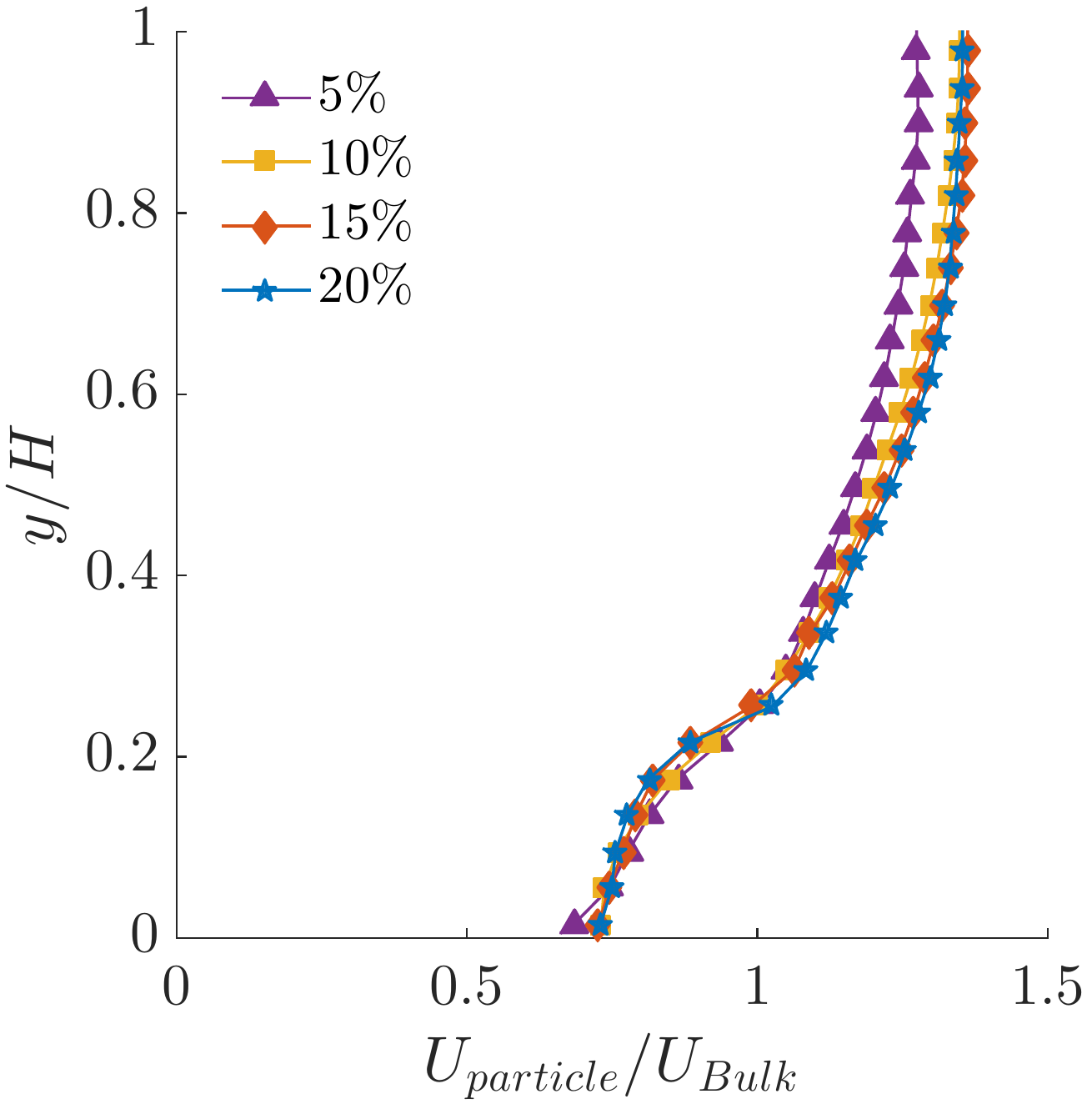}
  \caption{}
  \label{fig:U_particle_NN}
\end{subfigure}
\begin{subfigure}{.49\textwidth}
  \centering
  \includegraphics[height=1\linewidth]{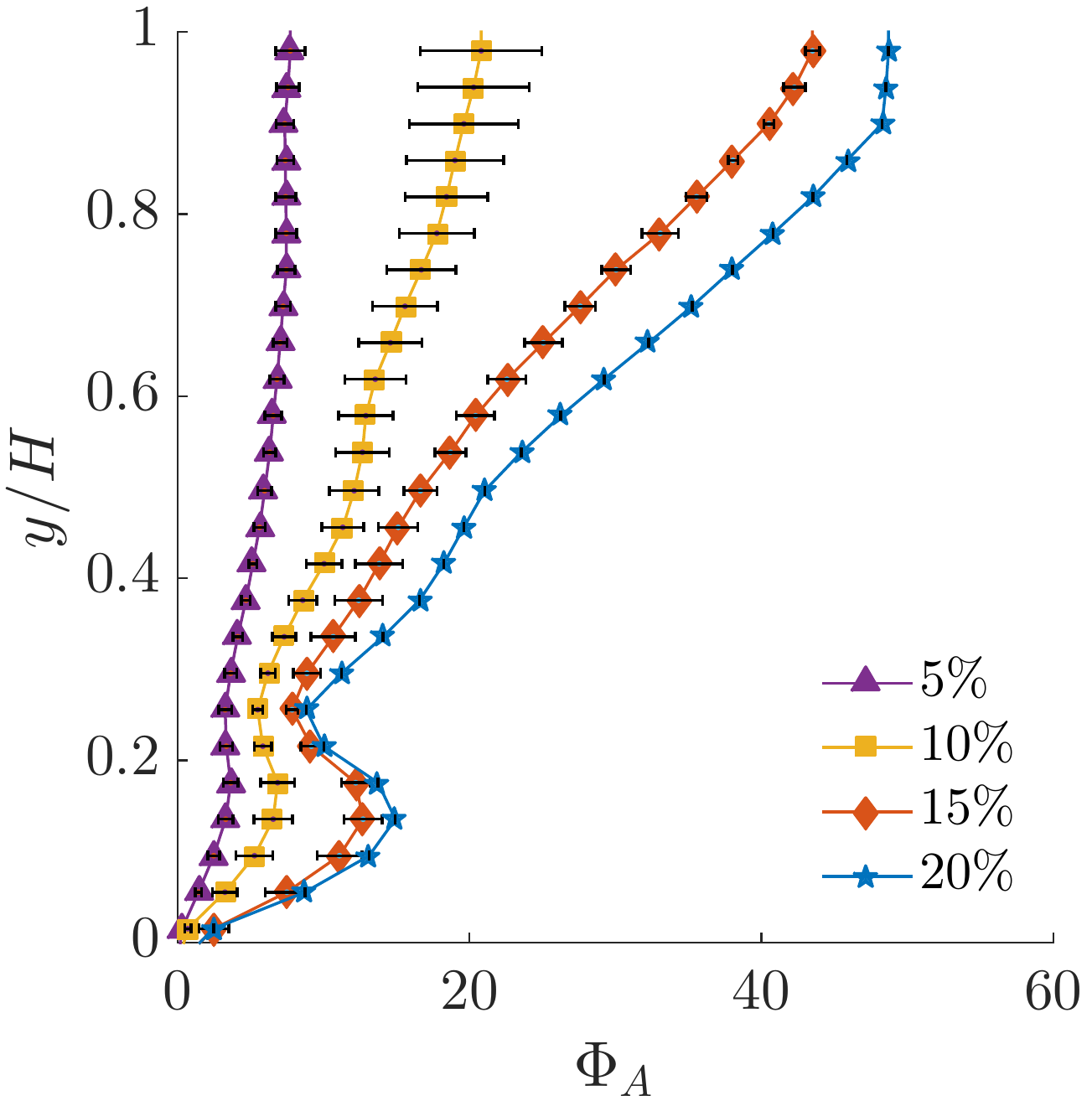}
  \caption{}
  \label{fig:Phi_NN}
\end{subfigure}
\caption{Particle-laden viscoelastic fluid flow: Mean stream-wise velocity profiles for (a) fluid phase and (b) particle phase. Panel (c) depicts the particle (area) concentration profile.}
\label{fig:U_mean_and_Phi_NN}
\end{figure}

Figure \ref{fig:U_fluid_NN} and \ref{fig:U_particle_NN} shows the mean streamwise velocity profiles for the fluid and particle phase in the plane of the wall-bisector $z/H$ = 0. The profile for single phase NF is also shown for comparison. As seen previously for NF, addition of particles makes the fluid velocity profile less flat i.e.\ the ratio $U_{Max}/U_{Bulk}$ increases. In contrast with the NF, particles at $\phi$ = 15\% result in the maximum $U_{Max}/U_{Bulk}$. The particle mean streamwise velocity $U_{particle}$ closely follows the fluid velocity $U_{fluid}$ except in the near-wall region as seen for the NF. The particle area concentration profile is shown in figure \ref{fig:Phi_NN} and similar to the Newtonian case, particles migrate towards the core and also form a layer of high concentration at the wall. However, the migration towards the core is substantially higher than in the NF and the migration towards the wall is marginally lower (cf figure \ref{fig:Phi}). This will be discussed later after looking at the fluctuating velocity statistics.

\begin{figure}
\centering
\begin{subfigure}{.49\textwidth}
  \centering
  \includegraphics[height=1\linewidth]{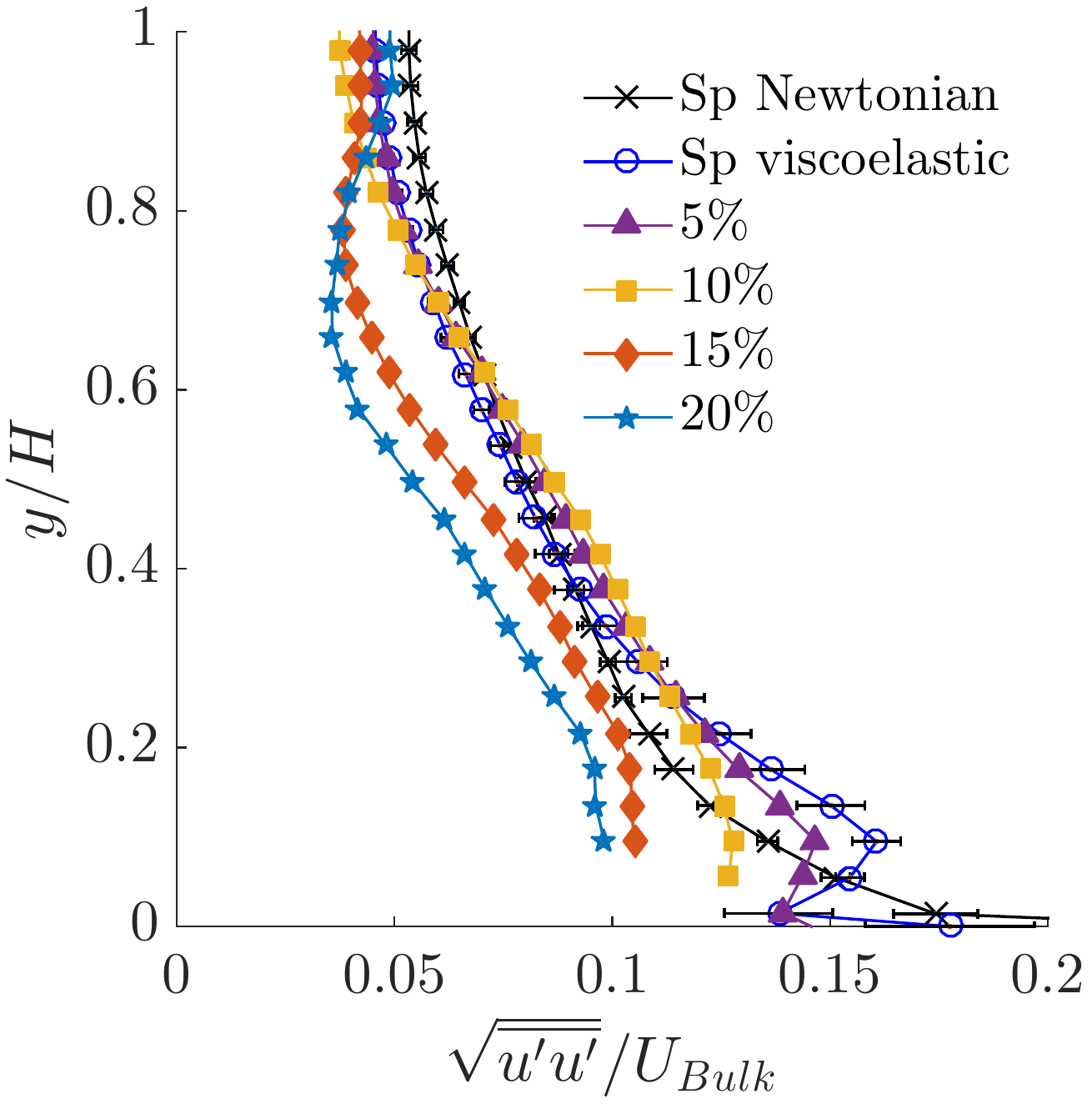}
  \caption{}
  \label{fig:urms_NN}
\end{subfigure}%
\begin{subfigure}{.49\textwidth}
  \centering
  \includegraphics[height=1\linewidth]{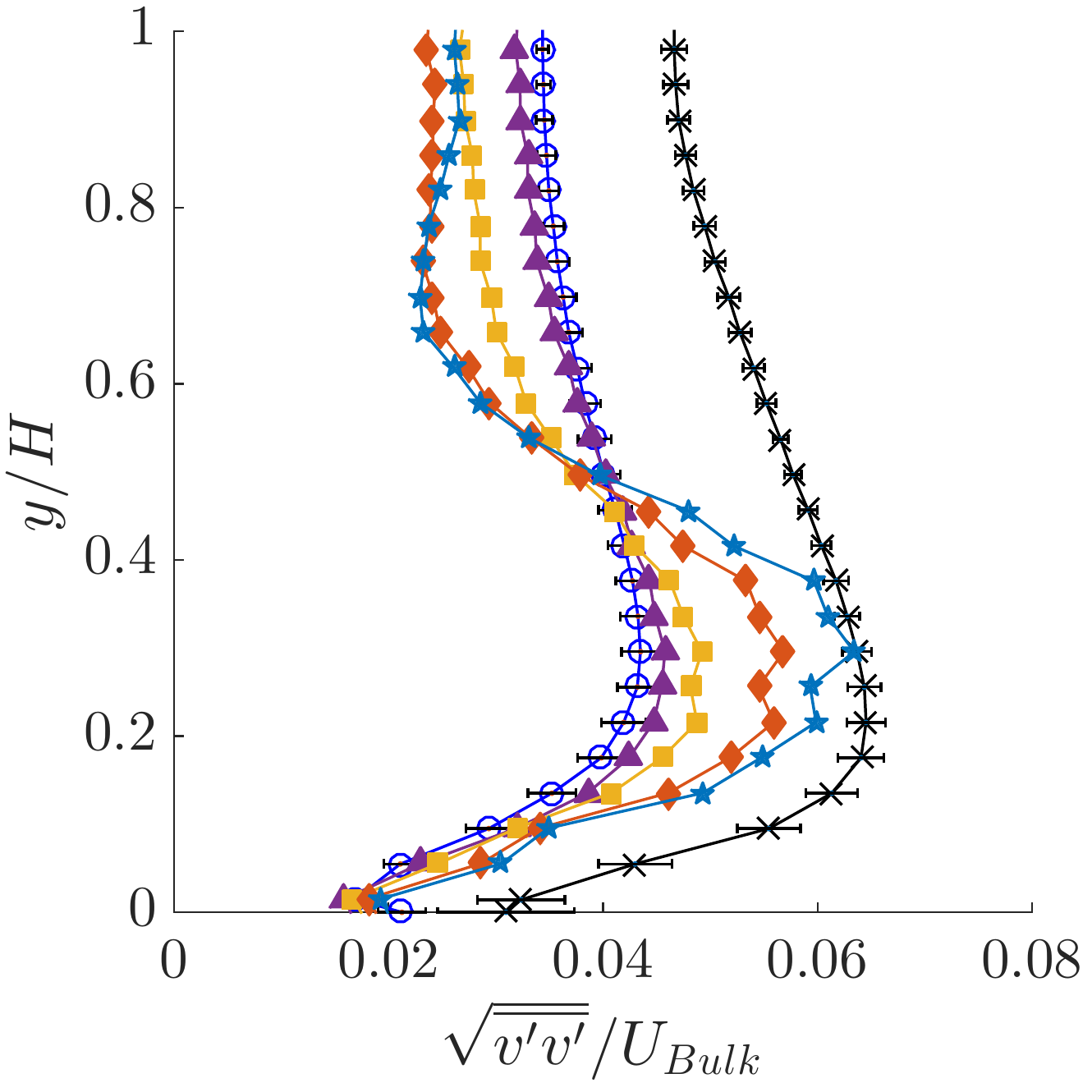}
  \caption{}
  \label{fig:vrms_NN}
\end{subfigure}
\begin{subfigure}{.49\textwidth}
  \centering
  \includegraphics[height=1\linewidth]{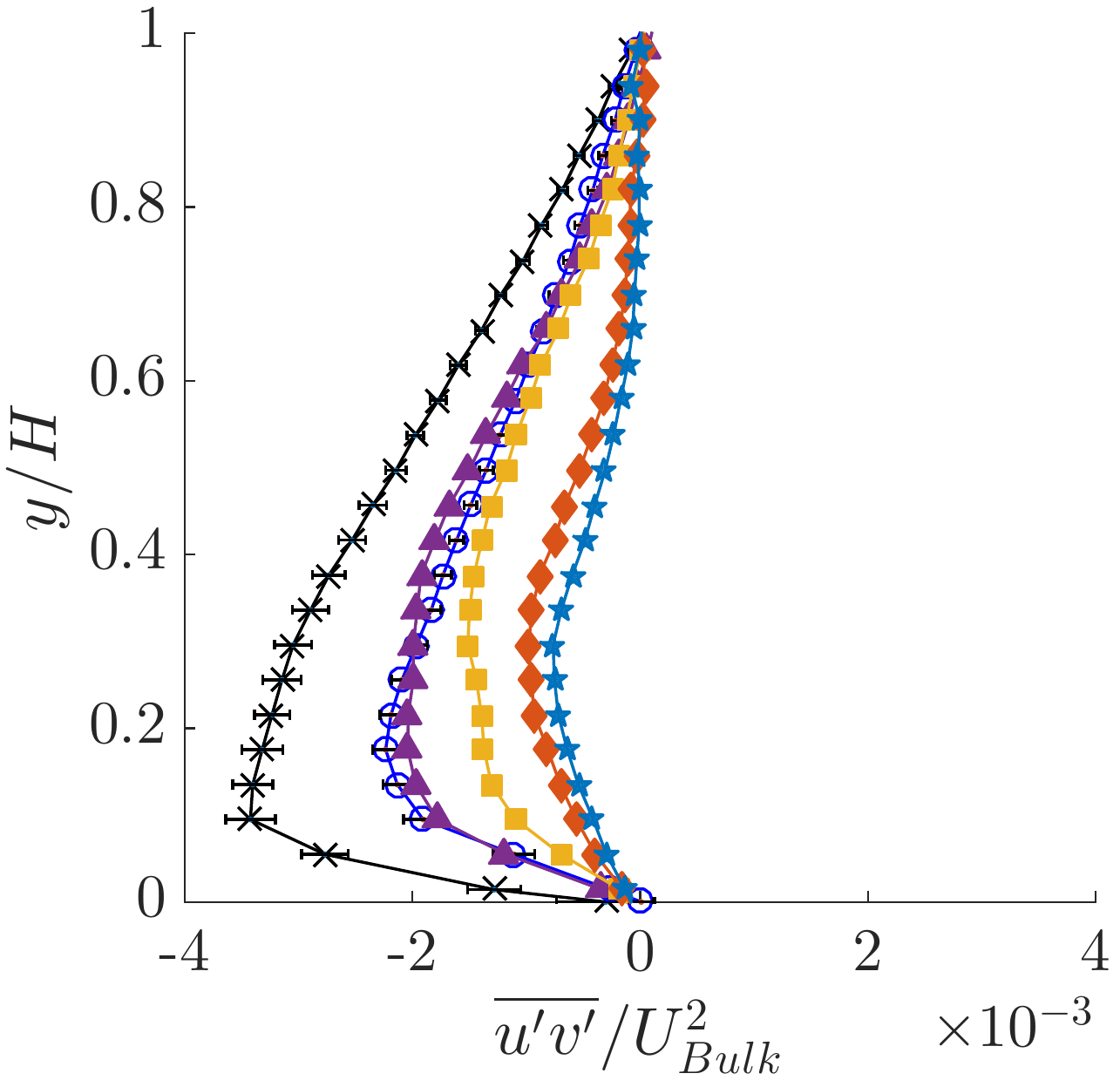}
  \caption{}
  \label{fig:uv_NN}
\end{subfigure}
\begin{subfigure}{.49\textwidth}
  \centering
  \includegraphics[height=1\linewidth]{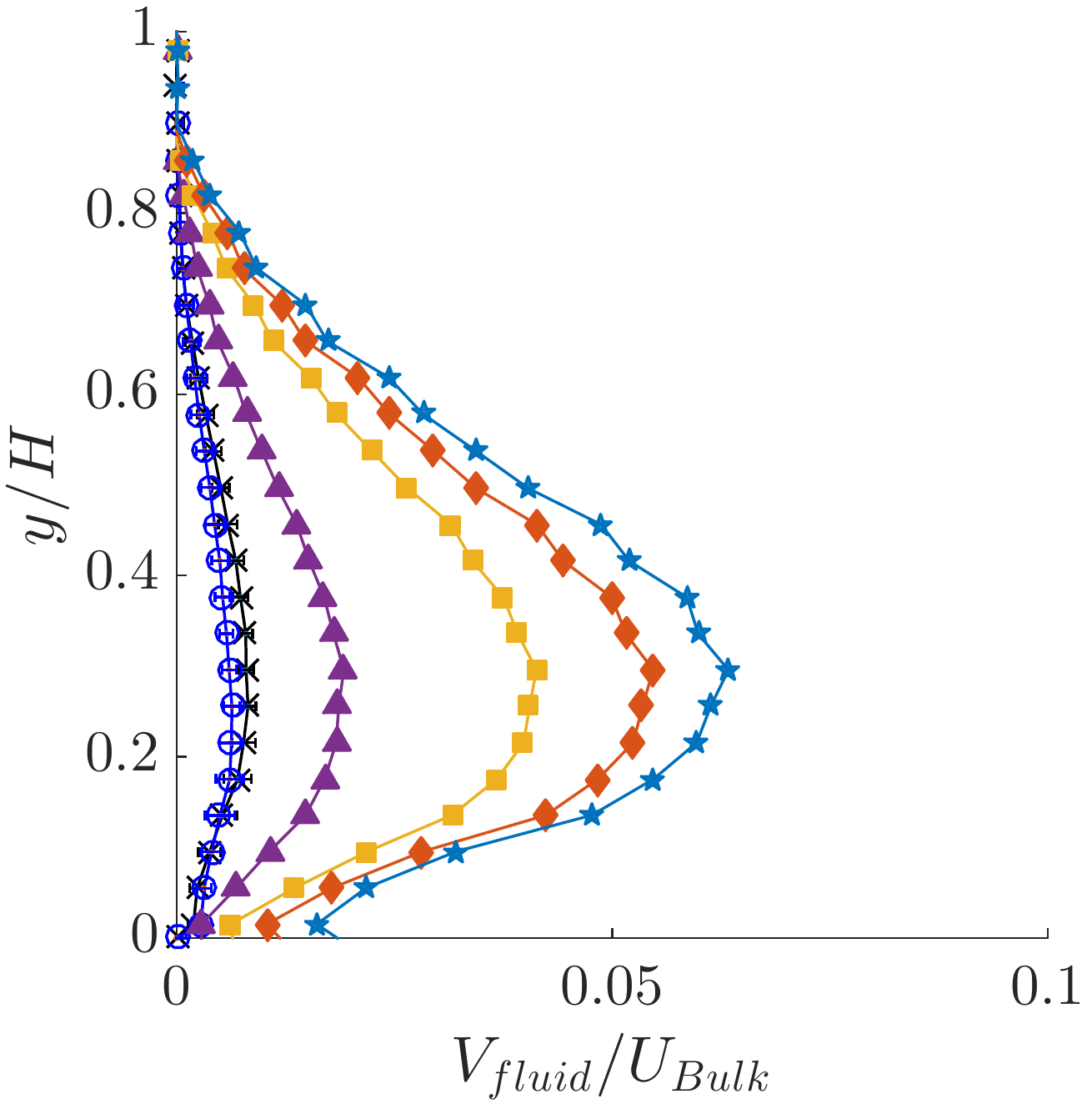}
  \caption{}
  \label{fig:V_NN}
\end{subfigure}
\caption{Particle-laden viscoelastic flow: Fluid velocity fluctuations in the (a) stream-wise direction and (b) wall-normal direction. Panels (c) shows the Reynolds shear stress and (d) shows the mean wall-normal velocity. Sp stands for Single phase.}
\label{fig:fluctuating_velocity_NN}
\end{figure}

Streamwise velocity fluctuations are shown in figure \ref{fig:urms_NN} together with the single phase NF is also shown. Similar to the NF case, the streamwise turbulence intensity, generally, decreases with increasing $\phi$. For $\phi$ = 5 and 10\%, there is a small increase between 0.2$\leq y/H \leq$0.6. At $\phi$ = 10\%, the peak value is damped and it becomes less distinct as compared to $\phi$ = 5\%. Values closest to the wall have not been plotted for higher $\phi$ = 10--20\% due to larger uncertainties in their measurement. Away from the wall, streamwise fluctuations reduce with increasing $\phi$. However, closer to the center, increasing $\phi$ from 10 to 20\% marginally increase the fluctuations, which is not observed for the Newtonian case.

The wall-normal fluctuations exhibit a peculiar behaviour compared to the Newtonian case. As shown in figure \ref{fig:vrms_NN}, these fluctuations progressively increase below $y/H\approx$ 0.4 and decrease above it. Fluctuations are stronger in the region of lower particle concentration (see figure \ref{fig:Phi_NN}). The peak value of these fluctuations increase with $\phi$ and for $\phi$ = 20\%, it is very similar to the NF cases (see figure \ref{fig:vrms}). Interestingly, profiles for $\phi$ = 20\% in both Newtonian and viscoelastic fluids appear quite similar suggesting that at high enough $\phi$, fluid wall-normal velocity fluctuations are dominated by the particle dynamics. \sz{In VEF flow, the transfer of energy from the streamwise to the wall-normal velocity fluctuations (through pressure-strain redistribution) is inhibited. This leads to reduced wall-normal fluctuations \cite{walker1990turbulent}. The increase in wall-normal fluctuations below $y/H\approx$ 0.4 in figure \ref{fig:vrms_NN} may be due to an enhanced particle-induced transfer of energy from the streamwise fluctuations.}

The primary Reynolds shear stress scaled by $U_{Bulk}^2$ also decreases with increasing $\phi$ for all $y/H$ except for the lowest $\phi$ = 5\%, where there is a small increase above $y/H$ = 0.3 as also seen in the NF case. For the highest $\phi$ = 20\%, the Reynolds shear stress reaches very small values indicating poor correlation between the streamwise and wall-normal fluctuations and their reduced contribution to the fluid momentum transport.

The mean secondary flow velocity seems to increase with $\phi$ as seen in figure \ref{fig:V_NN}. The increase is evidently more, almost two-times, than the corresponding NF case (cf figure \ref{fig:V}). The origin of this secondary motion in particle-laden duct flows, as stated before, depends on quantities which have not been measured and hence, it is difficult to speculate the reason behind this higher increase.

\section{Conclusion and discussion}

We have reported and discussed experimental results concerning velocity and particle concentration distribution in the plane of the wall-bisector of a square duct. The suspension consists of nearly neutrally-buoyant finite-sized spherical particles ($2H/d_p$ = 10) in a turbulent Newtonian (NF) and viscoelsatic fluid (VEF) flow at the same Reynolds number $Re_{2H}$. In NF, the wall-friction or total shear stress at the wall is an increasing function of particle concentration $\phi$. For $\phi \leq$ 10\%, the magnitude of the friction factor is in satisfactory agreement with the friction factor estimated using an effective suspension viscosity $\eta_e$. The measured value increases more rapidly with $\phi$ than the estimate using $\eta_e$ and thus, at $\phi$ = 35\%, experiments measure a drag increase of around 90\% whereas the drag estimated using $\eta_e$ is only around 50\% higher than single phase NF at the same $Re_{2H}$. This happens primarily beacuse the particles are not uniformly distributed and undergo preferential migration towards the core and the wall, resulting in a non-uniform equilibrium concentration profile as shown in figure \ref{fig:Phi}. As discussed in \citet{lashgari2014laminar}, for a suspenion in Newtonian fluid flow, the total shear stress at the wall is due to fluid viscous stresses, fluid+particle turbulence stresses and particle-induced stresses. A square duct flow is non-homogeneous in the two cross-stream directions and hence, the stress balance has to be preformed for the entire cross section to evaluate the contribution from each stress components to the total shear stress. Nevertheless, measurements of these components in the plane of the wall-bisector provides important insights in to the overall stress budget. From figure \ref{fig:fluctuating_velocity}, it appears that the fluid turbulence is increasingly damped with increasing $\phi$: the primary Reynolds shear stress is lowest at the highest $\phi$. Simulations from \cite{fornari2017suspensions} indicate that the particle turbulent shear stress is smaller than the fluid phase across the entire cross-section, almost for all $\phi$. Thus the contribution of turbulent stresses from both the phases is expected to decrease with $\phi$. Fluid viscous stresses are expected to change marginally since the mean fluid velocity profile is not drastically altered compared to the single-phase case. Also, at high Reynolds number, the contribution of viscous stresses to the overall shear stress is small. Hence, the substantial increase in the total shear stress with $\phi$ occurs despite a reduction in the turbulent stresses, and hence, it is attributed to higher particle-induced stresses. Especially near the core, for $\phi$ = 20\%, local particle concentration reaches values as high as 40\% and one can expect that the relative contribution of the particle-induced stresses is highest in this region. \sz{Particle-induced stresses are also high in the region of the particle wall layer where there is significant slip between the two phases, as also seen in \cite{lashgari2016channel}.} To note, the particle-induced stresses contains contributions from the hydrodynamic stresslet, particle acceleration, and inter-particle collision \citep{zhang2010physics}.

\begin{figure}
\centering
 \includegraphics[height=0.50\linewidth]{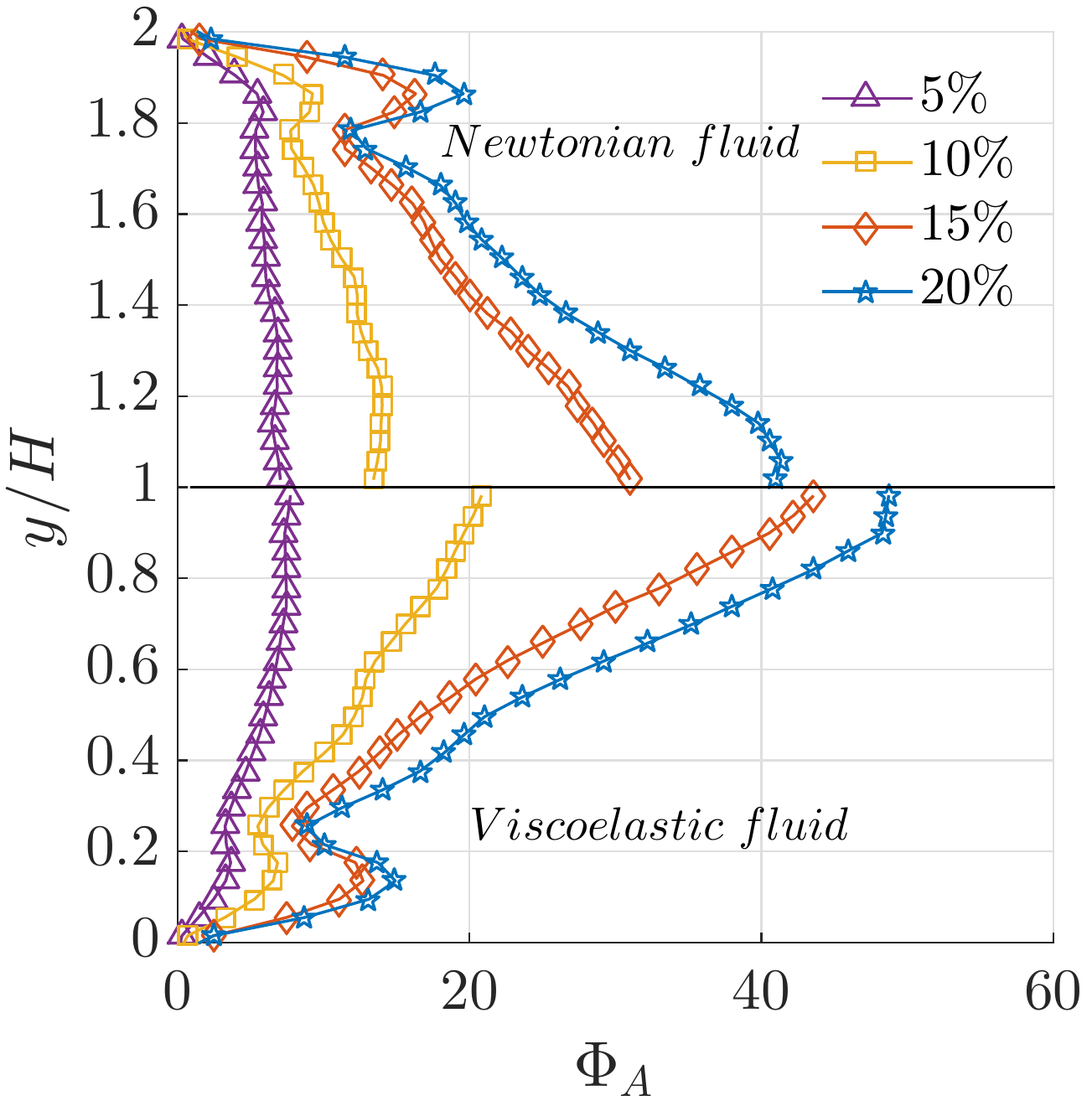}
  \caption{Particle concentration in the plane of the wall-bisector for particle-laden Newtonian and drag reducing fluid flow.}
  \label{fig:Phi_N_vs_NN}
\end{figure}

For the VEF, migration towards the core is more pronounced as can be seen in figure \ref{fig:Phi_N_vs_NN}, where the top half reports concentration profiles for the NF case and the bottom half for the VEF case. The relatively higher particle concentration in the core for the VEF case leads to a higher contribution of particle-induced stresses towards the total stress. This could explain why the rate of drag increase is higher for VEF as compared to NF as seen in figure \ref{fig:Phi_Drag_NN}. Even then, the absolute value of drag is still lower for particles in VEF than NF indicating that drag-reduction due to viscoelastic effects still influence the momentum transport, at least up to $\phi$ = 20\%. 

\begin{figure}
\centering
\begin{subfigure}{.49\textwidth}
  \centering
  \includegraphics[height=1\linewidth]{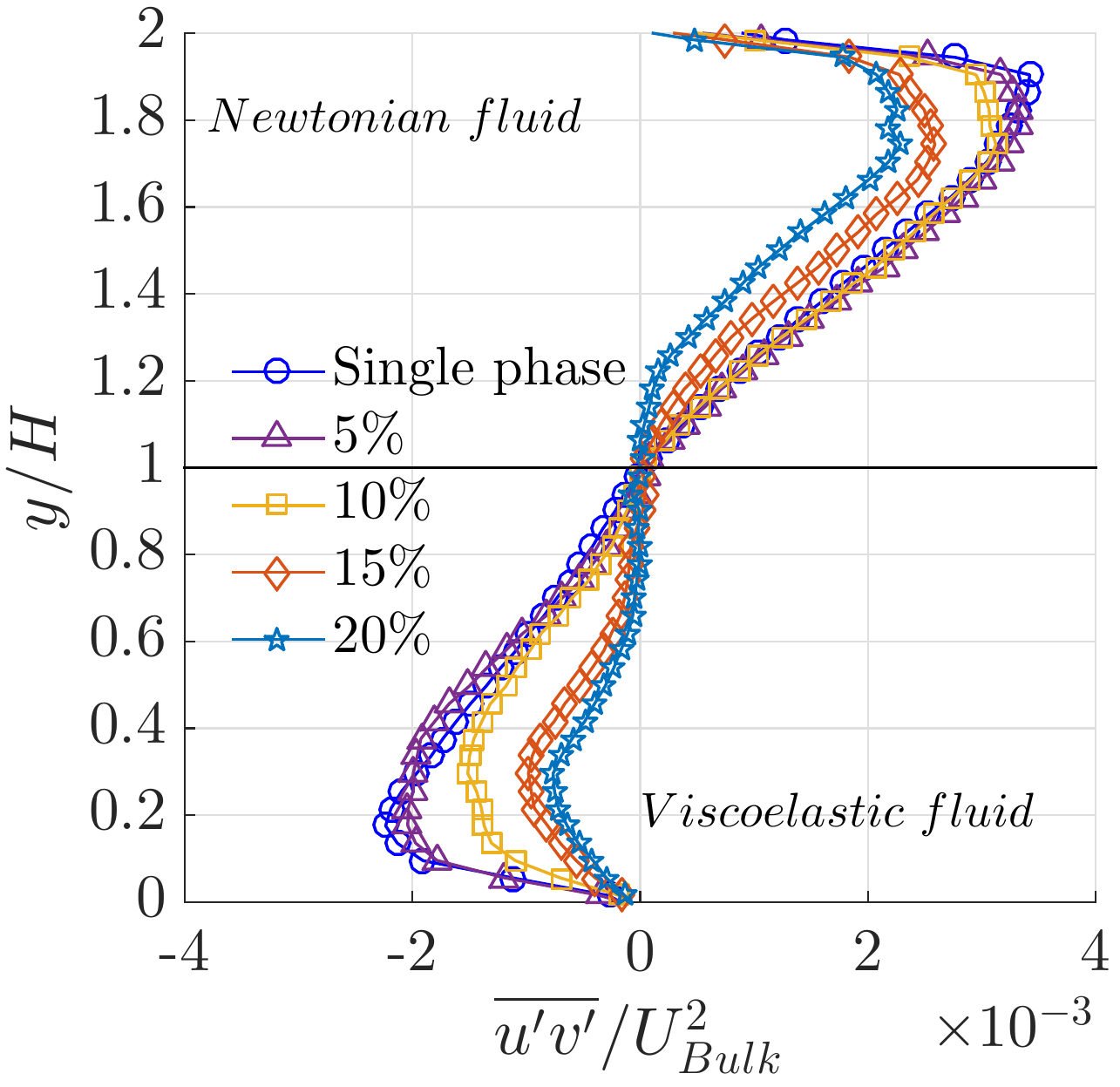}
  \caption{}
  \label{fig:uv_N_vs_NN}
\end{subfigure}
\begin{subfigure}{.49\textwidth}
  \centering
  \includegraphics[height=1\linewidth]{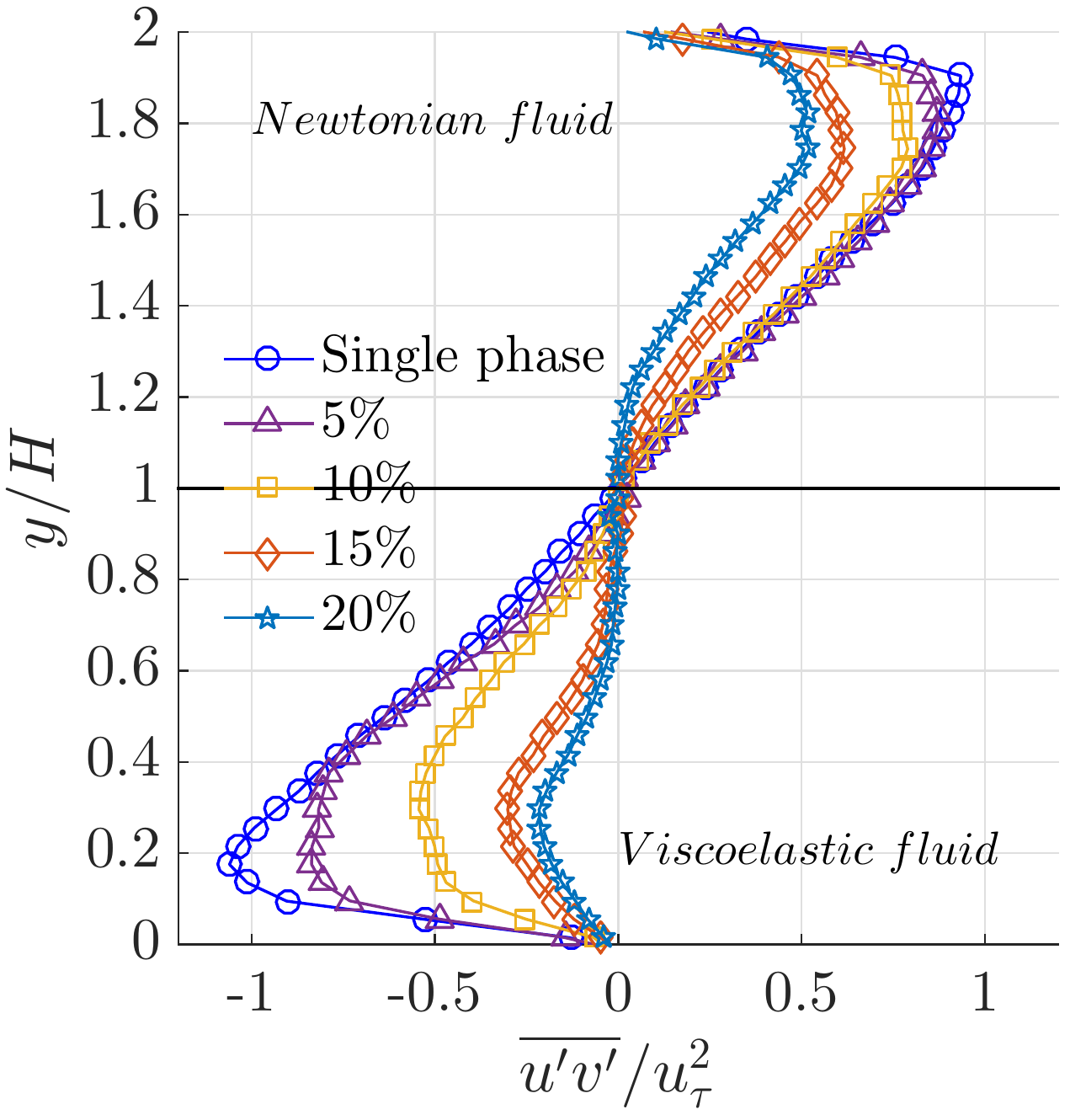}
  \caption{}
  \label{fig:uv_N_vs_NN_utau}
\end{subfigure}
\caption{Comparing Reynolds shear stress scaled using (a) $U_{Bulk}^2$ and (b) $u_{\tau}^2$ in the plane of the wall-bisector for particle-laden Newtonian and drag reducing fluid flow.} 
\label{fig:uv_N_vs_NN}
\end{figure}

Another interesting difference in the turbulence characteristics between VEF and NF can be seen from the profiles of the Reynolds shear stress compared in figure \ref{fig:uv_N_vs_NN}. When scaled by $U_{Bulk}^2$, the Reynolds shear stress for single phase VEF (lower half of figure \ref{fig:uv_N_vs_NN}) is lower than the corresponding NF (upper half of figure \ref{fig:uv_N_vs_NN}) and it reduces further with increasing particle concentration $\phi$. For the highest $\phi$ = 20\%, the turbulent shear stress in VEF is substantially smaller than the corresponding NF due to the combined action of particles and elasticity of the suspending media. Better understanding from a stress perspective can be obtained by plotting the Reynolds shear stress scaled by the average friction velocity $u_{\tau}^2$, as shown in figure \ref{fig:uv_N_vs_NN_utau}. Nearly for all $\phi$, the Reynolds shear stress is more suppressed for VEF than NF, indicating lower contribution of turbulent stresses to the total stress budget. Despite this larger suppression of turbulence, the rate of increase in total stress is higher for VEF due to the higher particle-induced stresses with $\phi$.

Finally, it may be speculated that at high enough $\phi$, turbulence would be highly suppressed, so that the turbulent stresses would have a negligible contribution to the total shear stress. In such a suspension, the flow would most likely be dominated by the particle dynamics as particles would become an increasingly important carrier of momentum. With reduction in turbulence and also reduced presence of the fluid phase in the mixture, drag-reduction due to polymer additives may become increasingly ineffective. \sz{The increase in drag at such high $\phi$ in VEF is certainly one aspect that remains to be seen. Future work should also study the effect of particle size on the flow statistics.}

\section*{Acknowledgements}

This work was supported by the European Research Council Grant No. ERC-2013-CoG-616186, TRITOS, from the Swedish Research Council (VR), through the Outstanding Young Researcher Award to LB. Åsa Engström (Rise Bioeconomy AB) is gratefully acknowledged for assistance with the rheological measurements.

\section*{References}
\bibliography{mybibfile}

\end{document}